\documentclass[aps,prb,twocolumn]{revtex4}
\usepackage{amssymb}
\usepackage{graphicx}
\usepackage{hyperref}

\begin{document}
\title{Superconducting and normal state properties of the layered boride OsB$_2$}
\author{Yogesh Singh, A. Niazi, M. D. Vannette, R. Prozorov, and D. C. Johnston}
\affiliation{Ames Laboratory and Department of Physics and Astronomy, Iowa State University, Ames, IA 50011}
\date{\today}

\begin{abstract}
OsB$_2$ crystallizes in an orthorhombic structure (\emph{Pmmn}) which contains alternate boron and osmium layers stacked along the \emph{c}-axis.  The boron layers consist of puckered hexagons as opposed to the flat graphite-like boron layers in MgB$_2$.  OsB$_2$ is reported to become superconducting below 2.1~K\@.  We report results of the dynamic and static magnetic susceptibility, electrical resistivity, Hall effect, heat capacity and penetration depth measurements on arc-melted polycrystalline samples of OsB$_2$ to characterize its superconducting and normal state properties.  These measurements confirmed that OsB$_2$ becomes a bulk superconductor below $T_{\rm c} = 2.1$~K\@.  Our results indicate that OsB$_2$ is a moderate-coupling Type-II superconductor with an electron-phonon coupling constant $\lambda_{\rm ep} \approx 0.4$--0.5, a small Ginzburg-Landau parameter $\kappa\sim$ 1--2 and an upper critical magnetic field $H_{\rm c2}$(0.5~K)~$\sim 420$~Oe for an unannealed sample and $H_{\rm c2}$(1~K)~$\sim 330$~Oe for an annealed sample.  The temperature dependence of the superfluid density $n_{\rm s}(T)$ for the unannealed sample is consistent with an \emph{s}-wave superconductor with a slightly enhanced zero temperature gap $\Delta(0) = 1.9~k_{\rm B}T_{\rm c}$ and a zero temperature London penetration depth $\lambda(0)$~=~0.38(2)~$\mu$m\@.  The $n_{\rm s}(T)$ data for the annealed sample shows deviations from the predictions of the single-band $s$-wave BCS model.  The magnetic, transport and thermal properties in the normal state of isostructural and isoelectronic RuB$_2$, which is reported to become superconducting below 1.6~K, are also reported. 
\end{abstract}

\maketitle

\section{Introduction}
\label{sec:INTRO}

Since the discovery of superconductivity in MgB$_2$ at a remarkably high temperature~$T_{\rm c} \approx 39$~K,\cite{Nagamatsu2001} there has been a renewed interest in the study of metal diborides. Many structurally-related \emph{T}B$_2$ compounds (\emph{T}~=~Ti, Zr, Hf, V, Cr, Nb, Ta, Mo) have been investigated in the search for superconductivity,\cite{Kaczorowski2001, Gasparov2001, Rosner2001} some of which had already been studied in the past.\cite{Cooper1970}  

Among all binary diborides with the AlB$_2$ structure, apart from MgB$_2$, superconductivity has only been reported for ZrB$_2$ ($T_{\rm c} \approx 5.5$~K),\cite{Gasparov2001} NbB$_2$ ($T_{\rm c}\approx 0.6$~K), Zr$_{0.13}$Mo$_{0.87}$B$_2$ ($T_{\rm c}\approx 5$~K) \cite{Cooper1970} and TaB$_2$ ($T_{\rm c}\approx 10$~K) \cite{Kaczorowski2001} although there are controversies about superconductivity in ZrB$_2$, NbB$_2$ and TaB$_2$. \cite{Gasparov2001, Rosner2001}  It has been argued using band structure calculations that in MgB$_2$, the high $T_{\rm c}$ is due to the B 2\emph{p} bands at the Fermi energy, and that any chemical, structural or other influence that changes this depresses $T_{\rm c}$. \cite{Medvedeva2001}  OsB$_2$ and RuB$_2$, which form in an orthorhombic structure (\emph{Pmmn}) containing deformed boron sheets instead of a planar boron array as in MgB$_2$, have also been reported to become superconducting below 2.1~K and 1.6~K, respectively.\cite{Vandenberg1975}  Recently the bulk modulus of OsB$_2$ at ambient and high pressure and its hardness have been studied. \cite{Cumberland2005}  Other physical properties of OsB$_2$ besides $T_{\rm c}$ have not yet been reported.  Band structure calculations suggest that OsB$_2$ and RuB$_2$ are indeed metallic. \cite{Hebbachea2006, Chen2006}  

\begin{figure}[t]
\includegraphics[width=3.25in]{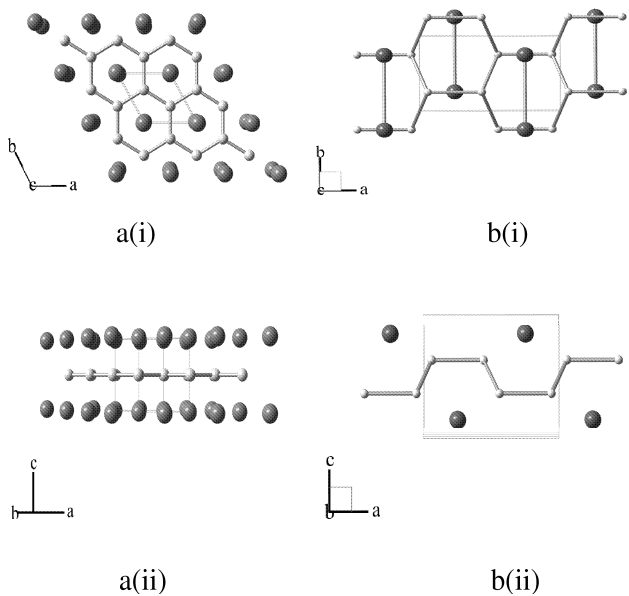}
\caption{The crystal structures of MgB$_2$ (a) and OsB$_2$ (b).  The transition metal atoms are shown as large spheres while the boron atoms are shown as the small spheres.  a(i) The MgB$_2$ structure viewed along the \emph{c}-axis and a(ii) perpendicular to the \emph{c}-axis .  The MgB$_2$ structure has alternate boron and transition metals planes stacked along the \emph{c}-axis.  The boron atoms form graphite-like sheets in the \emph{ab}-plane separated by a layer of transition metal atoms in a hexagonal close packing arrangement.\cite{Nagamatsu2001}~   b(i) The OsB$_2$ structure viewed along the \emph{c}-axis and b(ii) viewed perpendicular to the \emph{c}-axis.  The OsB$_2$ structure has a deformed two-dimensional network of corrugated boron hexagon sheets.  Along the c-axis the boron layers lie between two planar transition metal layers which are offset along the \emph{ab}-plane.\cite{Hebbachea2006}  
\label{Figstructure}}
\end{figure}

For comparison, the structures of MgB$_2$ and OsB$_2$ are shown in Fig.~\ref{Figstructure}.  While MgB$_2$ has flat graphite-like sheets of boron separated by a layer of transition metal atoms in a hexagonal close packing arrangement [Fig.~\ref{Figstructure}(a)],\cite{Nagamatsu2001} the OsB$_2$ structure has sheets of a deformed two-dimensional network of corrugated boron hexagons. The boron layers lie between two planar transition metal layers which are offset [Fig.~\ref{Figstructure}(b)].\cite{Hebbachea2006}

Herein we report the dynamic and static magnetic susceptibility, specific heat, resistivity, Hall effect and penetration depth studies on OsB$_2$ to characterize its superconducting and normal state properties.
We confirmed that OsB$_2$ is metallic and becomes superconducting below $T_{\rm c} = 2.1$~K\@.  Our results indicate that OsB$_2$ is a moderate-coupling superconductor with an electron-phonon coupling constant $\lambda_{\rm ep} \approx 0.4$--0.5, a small Ginzburg-Landau parameter $\kappa\sim$~1--2 and an upper critical magnetic field $H_{\rm c2}$(0.5~K) $\sim$ 420~Oe for an unannealed sample and $H_{\rm c2}$(1~K) $\sim$ 330~Oe for an annealed sample.  We also report measurements on RuB$_2$ which show similar normal state properties.  The paper is organized as follows.  Experimental details are given in Sec.~{\ref{sec:EXPT}}.  The structural results are presented in Sec.~{\ref{sec:RES-structure}}.  The normal state electrical resistivity, magnetic susceptibility and heat capacity data for the unannealed OsB$_2$ and RuB$_2$ samples are given in Sec.~{\ref{sec:normalstate}} and their superconducting properties are presented in Sec.~{\ref{sec:SC}}.  The normal state and superconducting properties of the annealed OsB$_2$ sample are presented in Sec.~(\ref{sec:annealed}).  The paper in concluded in Sec.~{\ref{sec:CON}}.

\section{EXPERIMENTAL DETAILS}
\label{sec:EXPT}
\noindent
The binary phase diagram of the Os-B system has recently been investigated in detail and it has been shown that OsB$_2$ melts congruently at about 1870~$^\circ$C. \cite{Stuparevic2004}  Polycrystalline samples ($\sim$ 1~g) of OsB$_2$ used in this study were therefore prepared by arc-melting.  The two OsB$_2$ samples for which the properties are reported here were prepared with starting materials of different purity.  One sample (sample~A) was prepared from Os powder (99.95\%, Alfa Aesar) and B chunks (99.5\%, Alfa Aesar).  The magnetization of this sample showed the presence of a large amount of paramagnetic impurities as apparent in the low temperature measurements.  Therefore, another sample of OsB$_2$ (sample~B) was prepared using ultrahigh purity Os (99.995\%, Sigma Aldrich) and $^{11}$B (99.999\%, Eagle Pitcher).  The magnetization for this sample showed that the concentration of paramagnetic impurities was considerably reduced compared to the first sample.  The superconducting transition temperature and transition width are similar for these two samples.  Most of the measurements pertaining to the superconducting state have been done on sample~A while sample~B has been used to obtain the intrinsic magnetic susceptibility of OsB$_2$ and for $^{11}$B NMR measurements which have been reported elsewhere.\cite{NMR}  

The samples were prepared as follows.  The constituent elements were taken in stoichiometric proportion and arc-melted on a water-cooled copper hearth in high purity argon atmosphere.  A Zr button was used as an oxygen getter.  The sample was flipped over and remelted 10-15 times to ensure homogeneous mixing of the constituent elements.  The mass of the ingot was checked after the initial two meltings and any weight loss due to the shattering of boron during melting was compensated by adding the appropriate amount of boron in subsequent melts.  The arc-melted ingot so obtained had a shiny metallic luster with well formed crystal facets on the surface.  A part of the sample~A was wrapped in Zr foil and annealed for 10 days at 1150~$^\circ$C in a sealed quartz tube.  A sample of the isostructural compound RuB$_2$ was prepared similarly from high purity Ru powder (99.995\%, MV labs) and $^{11}$B (99.999\%, Eagle Pitcher).  A portion of the as-cast samples was crushed for powder X-ray diffraction.  Powder X-ray diffraction (XRD) patterns were obtained using a Rigaku Geigerflex diffractometer with Cu K$\alpha$ radiation, in the 2$\theta$ range from 10 to 90$^\circ$ with a 0.02$^\circ$ step size. Intensity data were accumulated for 5~s per step.  

Samples of starting composition OsB$_{1.9}$ and OsB$_{2.1}$ were also prepared with the above starting materials from Alfa Aesar.  Powder X-ray diffraction measurements on crushed pieces of these samples showed that OsB$_{1.9}$ is a two-phase sample containing the phases Os$_2$B$_3$ and OsB$_2$ while the OsB$_{2.1}$ sample contained the phase OsB$_2$ and elemental osmium.  The purpose of making these samples was to explore both the boron-deficient and boron-rich sides, respectively, of the homogeneity range of OsB$_2$, if any, and its influence on the superconducting properties.

The temperature dependence of the dc magnetic susceptibility and isothermal magnetization was measured using
a commercial Superconducting Quantum Interference Device (SQUID) magnetometer (MPMS5, Quantum Design).  The resistivity and heat capacity were measured using a commercial Physical Property Measurement System (PPMS, Quantum Design).  The magnetic susceptibility measurements were done on samples of arbitrary shape.  The magnetization $M$ versus field $H$ measurements in Figs.~\ref{FigMH} and \ref{FigAN-OsB2-chi} below were performed on a parallelepiped sample with dimensions: length~=~3.25~mm, width~=~1.57~mm and thickness~=~0.35~mm with the magnetic field applied parallel to the length of the sample to minimize demagnetization effects.  The $M(H)$ loops in Figs.~\ref{Fig5quadMH} and \ref{FigAN-OsB2-chi} below were measured with the field applied either parallel and perpendicular to the length of the sample.  The resistance was measured using a four-probe technique with an ac current of 5~mA along the long axis of the rectangular bar, corresponding to a current density of 0.91~A/cm$^2$\@.  

The dynamic susceptibility was measured between 0.5~K and 2.5~K using a 10~MHz tunnel-diode driven oscillator (TDO) circuit with a volume susceptibility sensitivity $\Delta\chi\approx 10^{-8}$.\cite{prozorov2006r}  For superconductors, this is equivalent to a change in the London penetration depth of about 0.5~\AA ~for millimeter-sized samples.\cite{prozorov2006r,prozorov2000,prozorov2000a}  A TDO is an \emph{LC} tank circuit with a coil of inductance $L$ and a capacitor $C$. The circuit is self-resonating at a frequency $2\pi f=1/\sqrt{LC}$.  When a sample with susceptibility $\chi$ is inserted into the coil, the total inductance decreases for a diamagnetic sample or increases for a paramagnetic sample. The resonant frequency changes accordingly by an amount which is proportional to $\chi$.\cite{prozorov2006r}  Specifically, the device measures the temperature dependence of the resonant frequency shift $\Delta f(T)$ induced by changes in the sample's magnetic response.  The magnetic susceptibility $\chi$ is then given by\cite{prozorov2000a} 

\begin{equation}
\Delta f(T) = -4\pi\chi(T)G \approx -G\Big[1-\frac{\lambda\left(  T\right)  }{R}%
\tanh\left(  \frac{R}{\lambda\left(  T\right)  }\right)\Big]
\label{resonatorresponse}
\end{equation}
where \emph{G} is a sample-shape and coil-dependent calibration parameter, $R$ is the effective sample dimension and $\lambda(T)$ is the London penetration depth. 
The $G$ has been determined by matching the temperature dependence of the skin depth obtained from the resonator response in the normal state of OsB$_2$ to the measured resistivity data.

\section{RESULTS}
\subsection{Structures of OsB$_2$ and RuB$_2$}
\label{sec:RES-structure}

\begin{figure}[t]
\includegraphics[width=3.5in]{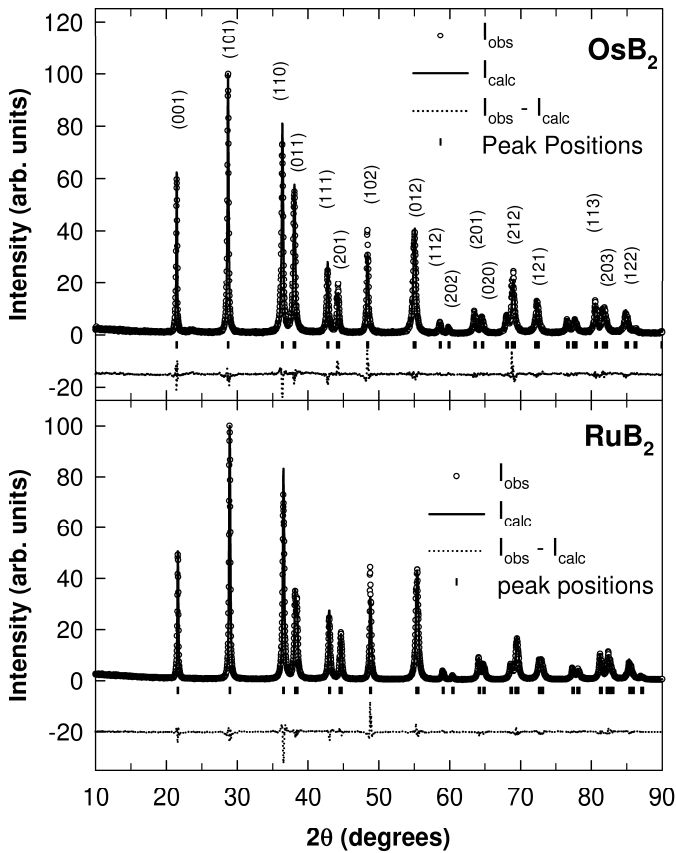}
\caption{Rietveld refinements of the OsB$_2$ and RuB$_2$ X-ray diffraction data. The open symbols represent the observed X-ray pattern, the solid lines represent the fitted pattern, the dotted lines represent the difference between the observed and calculated intensities and the vertical bars represent the peak positions.  
\label{Figxrd}}
\end{figure}

\begin{table*}

\caption{\label{tabStruct}
Structure parameters for OsB$_2$ and RuB$_2$ refined from powder
XRD data.  $B_{11}$, $B_{22}$ and $B_{33}$ are the anisotropic thermal parameters defined within the thermal parameter of the intensity as $e^{-(B_{11}h^2+B_{22}k^2+B_{33}l^2)}$.}
\begin{ruledtabular}
\begin{tabular}{l|c||ccccccc}
Sample & atom & \emph{x} & \emph{y} & \emph{z} & \emph{B}$_{11}$ & \emph{B}$_{22}$ & \emph{B}$_{33}$ & $R_{\rm wp}/R_{\rm p}$\\
  &  & & && (\AA$^2$) & (\AA$^2$) & (\AA$^2$) & \\\hline  
OsB$_2$ & Os~~ & 0.25 & 0.25 & 0.1515(3) & 0.0209(9) & 0.045(2) &	0.001(1)& 1.39\\  
& B~~ &0.093(3) & 0.25 & 0.669(4) &  & & &  \\\hline  
RuB$_2$ & Ru~~ &0.25 & 0.25 &	0.1490(2) &	0.0166(7) & 0.051(2) &0.0007(9)&  1.36\\
&B ~~& 0.063(1)& 0.25 & 0.638(2) &	 &  & &  \\
\end{tabular}
\end{ruledtabular}
\end{table*}

All the lines in the X-ray patterns for OsB$_2$ and RuB$_2$ could be indexed to the known orthorhombic \emph{Pmmn} (No.~59) structure and Rietveld refinements,\cite{Rietveld} shown in Fig.~\ref{Figxrd}, of the X-ray patterns gave the lattice parameters \emph{a}~=~4.6851(6)~\AA, \emph{b}~=~2.8734(4)~\AA~and \emph{c}~=~4.0771(5)~\AA~ for OsB$_2$ and \emph{a}~=~4.6457(5)~\AA, \emph{b}~=~2.8657(3)~\AA~and \emph{c}~=~4.0462(4)~\AA~ for RuB$_2$.  These values are in excellent agreement with previously reported values.\cite{Roof1962}  The best fits were obtained when the anisotropic thermal parameters for the transition metal atom were allowed to vary. For boron the overall isotropic thermal parameter was fixed to zero because unphysically large values were obtained when it was allowed to vary and fixing it to zero did not change the quality of the fit.  This is probably because the atomic number of boron is much less than that of either Os or Ru.  A neutron diffraction study is needed to obtain reliable estimates of the thermal parameters for boron.  Some parameters obtained from the Rietveld refinement are given in Table~\ref{tabStruct}.  
Although the lattice parameters and fractional atomic positions that we obtain from the Rietveld refinements for both OsB$_2$ and RuB$_2$ agree reasonably well with the earlier structural report,\cite{Roof1962} the fits consistently underestimate the intensities of the (102) peaks at 2$\theta~=~48.35^\circ$ (see Fig.~\ref{Figxrd}).

\subsection{Unannealed Samples}
\subsubsection{Normal State Properties of OsB$_2$ and RuB$_2$}
\label{sec:normalstate}

\label{sec:RES-normalstate-resistivity}
The electrical resistivity ($\rho$) versus temperature of OsB$_2$ (sample~A) and RuB$_2$ from 1.75~K to 300~K is shown in Fig.~\ref{Fig(Os,Ru)B2RES}.  The room temperature resistivity values are $36(3)~\mu$$\Omega$~cm for OsB$_2$ and $53(5)~\mu$$\Omega$~cm for RuB$_2$.  The error in the resistivity comes primarily from the error in the determination of the geometrical factors.  Both compounds show metallic behavior with an approximately linear decrease in resistivity on cooling from room temperature.  At low temperatures $\rho$ becomes only weakly temperature dependent and reaches a residual resistivity $\rho_0$ of $1.7(2)~\mu$$\Omega$~cm just above 2.2~K\@ for OsB$_2$ and $1.1(1)~\mu$$\Omega$~cm at 1.8~K for RuB$_2$ as seen in the inset of Fig.~\ref{Fig(Os,Ru)B2RES}.  The large residual resistivity ratios RRR~=~$\rho$(300~K)/$\rho_0$~=~22 for OsB$_2$ and RRR~=~51 for RuB$_2$ indicate well-crystallized homogeneous samples.  For OsB$_2$ the resistivity drops abruptly below 2.2~K and reaches zero by 2.14~K, as highlighted in the inset of Fig.~\ref{Fig(Os,Ru)B2RES}.  The superconducting properties will be discussed in detail in Sec.~{\ref{sec:SC}}. 
   
\begin{figure}[t]
\includegraphics[width=3in]{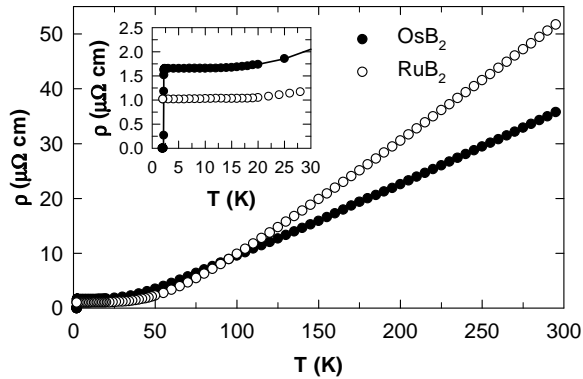}
\caption{Electrical resistivity $\rho$ for OsB$_2$ and RuB$_2$ versus temperature \emph{T}. The inset shows the low temperature data on an expanded scale to highlight the low residual resistivity. 
\label{Fig(Os,Ru)B2RES}}
\end{figure}

\label{sec:RES-normalstate-susceptibility}
\begin{figure}[t]
\includegraphics[width=3in]{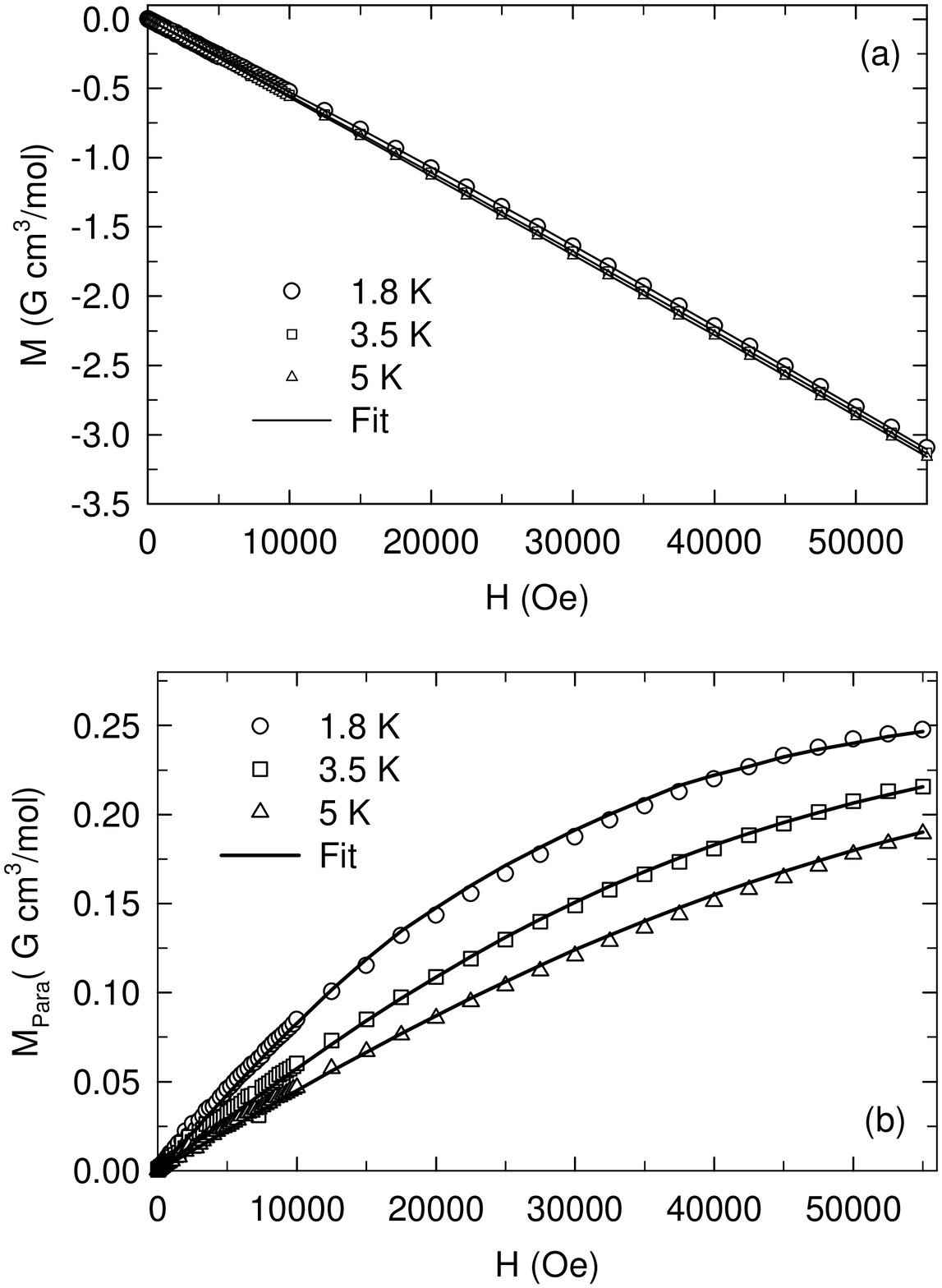}
\caption{(a) The magnetization $M$ versus applied magnetic field $H$ for OsB$_2$ (sample B).  The solid curves are fits by Eq.~(\ref{EqBF}).  (b) The paramagnetic impurity contribution $M_{\rm Para}$ to the total $M(H)$. The solid curves are the paramagnetic impurity part of Eq.~(\ref{EqBF}). 
\label{Figimpcontr}}
\end{figure}

Both OsB$_2$ and RuB$_2$ are so weakly magnetic that even trace amounts (few ppm) of magnetic impurities are apparent in the low temperature susceptibility and magnetization {$M(H)$} measurements.  The ${M(H)}$ data for OsB$_2$ at 1.8~K, 3.5~K and 5~K are shown in Fig.~\ref{Figimpcontr}(a).  The data were fitted by the expression 

\begin{equation}
M(H) = \chi_0H + fN_{\rm A}gS\mu_{\rm B}B_S(x) \equiv \chi_0H + M_{\rm Para}(H)~,
\label{EqBF}
\end{equation}
\noindent
where $\chi_0$ is the intrinsic susceptibility of OsB$_2$, \emph{f} is the molar fraction of paramagnetic impurities, $N_{\rm A}$ is Avogadro's number, \emph{g} is the \emph{g}-factor of the impurity spins, \emph{S} is the spin of the paramagnetic impurities and $B_S(x)$ is the Brillouin function where the argument \emph{x} of the Brillouin function is $x = g\mu_{\rm B}SH/(T-\theta)$ where $\theta$ is the Weiss temperature.  The \emph{g} value was fixed to two.  The fitting parameters were $\chi_0$, \emph{f}, \emph{S} and $\theta$ and we obtained $\chi_0 = -6.09(1)\times10^{-5}$~cm$^3$/mol, $f = 1.57(4)\times10^{-5}$, $S = 1.54(6)$ and $\theta = -1.8(2)$~K\@.  The fit is shown as the solid curves in Fig.~\ref{Figimpcontr}(a).  
By subtracting from the observed ${M(H)}$ data the $\chi_0 H$ obtained from the fitting, one can obtain the contribution from paramagnetic impurities $M_{\rm Para}$ as shown in Fig.~\ref{Figimpcontr}(b).  The solid curves in Fig.~\ref{Figimpcontr}(b) are the paramagnetic impurity part of the fit [the second term in Eq.~(\ref{EqBF})].  

\begin{figure}[t]
\includegraphics[width=3in]{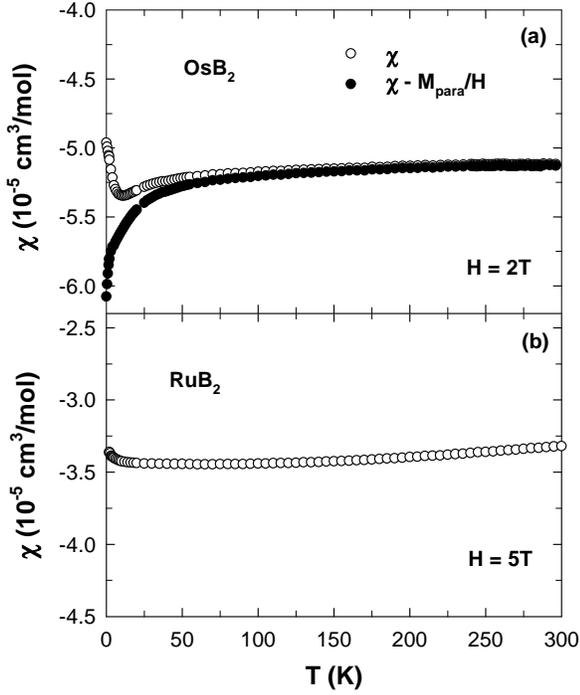}
\caption{Magnetic susceptibility $\chi$ (open circles) versus temperature \emph{T} for OsB$_2$ (a) and RuB$_2$ (b).  In (a), the filled circles are the values obtained after subtracting the paramagnetic impurity contribution (see text). 
\label{Figchi_normal}}
\end{figure}

The normal state susceptibility $\chi\equiv M/H$ for OsB$_2$ and RuB$_2$ has been measured versus temperature \emph{T} between 1.8~K and 300~K in a field of 2~T and 5~T, respectively, as shown in Fig.~\ref{Figchi_normal} (the susceptibility of the sample holder was corrected for).  The $\chi(T)$ for both samples is weakly temperature dependent between 50~K and 300~K\@. For OsB$_2$, below 50~K $\chi(T)$ drops somewhat on cooling, amounting to about 4\% of the room temperature value. The upturn at low temperatures seen for both samples is most likely due to the presence of small amounts (a few ppm) of paramagnetic impurities as determined above for OsB$_2$ (\emph{f}~=~16 molar~ppm).  Figure~\ref{Figchi_normal}(a) also shows the susceptibility of OsB$_2$ after subtracting the paramagnetic impurity contribution $\chi_{\rm Para} = M_{\rm Para}/H$ from the observed $\chi$.

The intrinsic susceptibility after subtracting $\chi_{\rm Para}$, $\chi(T)$, can be written as 

\begin{equation}
\chi~=~\chi_{\rm core}+\chi_{\rm L}+\chi_{\rm VV}+\chi_{\rm P}~,
\label{Eqcontributionstochi}
\end{equation}
\noindent
where $\chi_{\rm core}$ is the diamagnetic orbital contribution from the electrons (ionic or atomic), $\chi_{\rm L}$ is the Landau orbital diamagnetism of the conduction electrons, $\chi_{\rm VV}$ is the Van Vleck paramagnetic orbital contribution and $\chi_{\rm P}$ is the Pauli paramagnetic spin susceptibility of the conduction electrons. For OsB$_2$ and RuB$_2$, the net diamagnetic susceptibility indicates quasi-free electrons with $\chi_{\rm L}=-({m\over m^*})^2{\chi_{\rm P}\over 3}$ and $\chi_{\rm VV}\approx 0$,\cite{Peierls1955} where \emph{m} is the free electron mass and $m^*$ is the effective mass of the current carriers.  Assuming $m^* = m$, the Pauli susceptibility can be written as
\begin{equation} 
\chi_{\rm P}={3\over 2}(\chi-\chi_{\rm core})~,
\label{EqchiPauli}
\end{equation}

\noindent 
which can be extracted from the experimentally measured susceptibility, after correction for the paramagnetic impurity contribution, if the contribution from the core $\chi_{\rm core}$ is known. In covalent metals it is difficult to correctly estimate $\chi_{\rm core}$ because its value depends on the local electron density on the atoms. In an ionic model, $\chi_{\rm core}=-18\times 10^{-6}~{\rm cm^3/mol}$ for Os$^{6+}$ and $-44\times 10^{-6}~{\rm cm^3/mol}$ for Os$^{2+}$.\cite{Shannon1976}  However if we use the ionic core diamagnetism values for Os and B, one obtains from Eq.~(\ref{EqchiPauli}) an unphysical negative $\chi_{\rm P}$. For OsB$_2$ it is reasonable to use atomic (covalent) estimates of $\chi_{\rm core}$ instead of ionic values because the bonding in OsB$_2$ has a strong covalent character.\cite{Chiodo2006}  Therefore using the atomic diamagnetism values $\chi_{\rm core}$ for Os ($-53.82\times 10^{-6}~{\rm cm^3/mol}$), Ru ($-42.89\times 10^{-6}~{\rm cm^3/mol}$) and B ($-12.55\times 10^{-6}~{\rm cm^3/mol}$),\cite{Mendelsohn1970} we obtain $\chi_{\rm core}=-78\times10^{-6}~{\rm cm^3/mol}$ for OsB$_2$ and $\chi_{\rm core}=-68\times10^{-6}~{\rm cm^3/mol}$ for RuB$_2$.  Subtracting these values and the paramagnetic impurity contribution from the total measured susceptibility and accounting for the Landau diamagnetism, one can get the Pauli paramagnetic susceptibility using Eq.~(\ref{EqchiPauli}), thus yielding $\chi_{\rm P}$~=~2.7(3)$\times 10^{-5}$ cm$^3$/mol at $T$~=~0~K for OsB$_2$. 
 
From $\chi_{\rm P}$ one can estimate the density of states at the Fermi level $N(\epsilon_{\rm F})$ for both spin directions using the relation 
\begin{equation}
\chi_{\rm P}~=~\mu_{\rm B}^2 N(\epsilon_{\rm F})~, 
\label{EqDOSCHIP}
\end{equation}
\noindent
where $\mu_{\rm B}$ is the Bohr magneton.  Taking the above $T$~=~0~K value of $\chi_{\rm P}$ for OsB$_2$, we get $N(\epsilon_{\rm F})$~=~$1.7(2)~$states/(eV~f.u.), where ``f.u.'' means ``formula unit''.  This value is larger than the value from band structure calculations [$N(\epsilon_{\rm F})\approx 0.55$~states/(eV~f.u.)].\cite{Hebbachea2006}  Similarly, from the average value of $\chi_{\rm P}$~=~$5.22(7)\times 10^{-5}~{\rm cm^3/mol}$ for RuB$_2$ one estimates using Eq.~(\ref{EqDOSCHIP}), $N(\epsilon_{\rm F})$~=~$3.0(2)~$states/(eV~f.u.).  This value is also larger than the value obtained from the band structure calculations [$N(\epsilon_{\rm F})\approx 0.53~$states/(eV~f.u.)].\cite{Hebbachea2006}  The discrepancy between experiment and theory for OsB$_2$ is discussed later in terms of the Stoner enhancement factor of the susceptibility.    

\label{sec:RES-normalstate-heatcap}
\begin{figure}[t]
\includegraphics[width=3in]{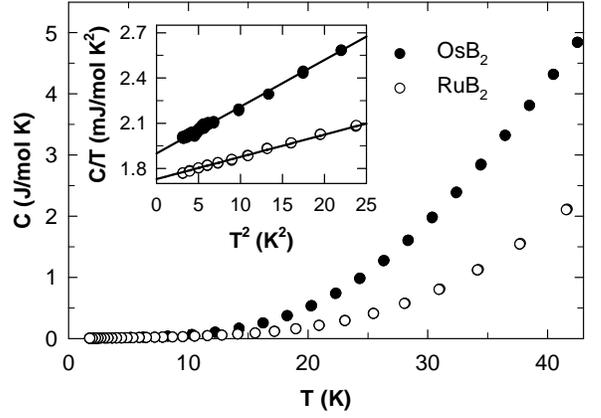}
\caption{The heat capacity \emph{C} versus temperature \emph{T} of OsB$_2$ and RuB$_2$ between 1.8~K and 45~K. The inset shows the data plotted as \emph{C/T} versus $T^2$ between 1.8~K and 5~K. The solid straight lines are fits by the expression $C/T~=~\gamma+\beta T^2$\@. 
\label{Fig(Os,Ru)B2Cp}}
\end{figure}

Figure~\ref{Fig(Os,Ru)B2Cp} shows the results of the normal state heat capacity versus temperature $C(T)$ measurements on OsB$_2$ (sample~A) and RuB$_2$, plotted as \emph{C} versus \emph{T}.  The OsB$_2$ data were recorded in an applied magnetic field of 1~kOe to suppress the superconducting transition to below 1.75~K, which is the low-temperature limit of our measurements.  The inset in Fig.~\ref{Fig(Os,Ru)B2Cp} shows the low temperature data for both samples plotted as \emph{C/T} versus $T^2$.  The low temperature data (1.75~K to 5~K) for both samples could be fitted by the expression $C~=~\gamma T+\beta T^3$ where the first term is the contribution from the conduction electrons and the second term is the contribution from the lattice.  The fits are shown as the solid straight lines in the inset of Fig.~\ref{Fig(Os,Ru)B2Cp}.  The values 
\begin{equation}
\gamma~=~1.90(1)~ {\rm mJ/mol~K}^2 ~~~{\rm and}~~~ \beta~=~0.031(2)~{\rm mJ/mol~K}^4
\label{OsB2gamma}
\end{equation}
are obtained for OsB$_2$ and the values 
\begin{equation}
\gamma~=~1.72(3)~ {\rm mJ/mol~K}^2 ~~~{\rm and}~~~ \beta~=~0.015(1)~{\rm mJ/mol~K}^4
\label{RuB2gamma}
\end{equation}
are obtained for RuB$_2$.  From the values of $\beta$, one can obtain the Debye temperature $\Theta_{\rm D}$ using the expression \cite{Kittel} 
\begin{equation}
\Theta_{\rm D}~=~\bigg({12\pi^4Rn \over 5\beta}\bigg)^{1/3}~, 
\label{EqDebyetemp}
\end{equation}
\noindent
where $R$ is the molar gas constant and $n$ is the number of atoms per formula unit (\emph{n}~=~3 for OsB$_2$ and RuB$_2$).  We obtain $\Theta_{\rm D}$~=~550(11)~K for OsB$_2$ and $\Theta_{\rm D}$~=~701(14)~K for RuB$_2$.  A simple harmonic oscillator model predicts $\Theta_{\rm D} \propto ({1\over M})^{1\over 2}$ (Ref.~\onlinecite{Kittel}) where \emph{M} is the molar mass of the compound.  The ratio ${\Theta_{\rm D}(OsB_2)\over \Theta_{\rm D}(RuB_2)} = 0.79(3)$ is indeed close to the square root of the ratio of the molar masses $\sqrt{M(RuB_2)\over M(OsB_2)}$~=~0.76.

Another quantity which characterizes a metal is the ratio of the density of states as probed by magnetic measurements to the density of states probed by heat capacity measurements, which is the Wilson ratio 
\begin{equation}
R_{\rm W}~=~{\pi^2 k_{\rm B}^2 \over 3\mu_{\rm B}^2}\bigg({\chi_{\rm P} \over \gamma}\bigg)~.
\label{EqWilson}
\end{equation}
\noindent
For a free-electron Fermi gas $R_{\rm W}$~=~1\@. Using $\chi_{\rm P}$~=~$2.7(3)\times 10^{-5}~{\rm cm^3/mol}$ and $\gamma$~=~$1.90(1)~{\rm mJ/mol}~{\rm K}^2$ for OsB$_2$ we get $R_{\rm W}$~=~1.03(2) at $T$~=~0~K which is of the order of unity expected for a quasi-free electron gas.  For RuB$_2$, $\chi_{\rm P}$~=~$5.22(7)\times 10^{-5}~{\rm cm^3/mol}$ and $\gamma$~=~$1.72(3)~{\rm mJ/mol}~K^2$ which gives $R_{\rm W}$~=~2.21(7)\@.  Note that Eq.~(\ref{EqWilson}) is only valid for spin-1/2 particles if the electron-phonon interaction and the Stoner enhancement factor are negligible.  If they are not, then one must use Eq.~(\ref{EqWilson1}) below.

\subsubsection{Superconducting Properties of OsB$_2$}
\label{sec:SC}
 \label{sec:RES-SC-resistivity-heatcap}
\begin{figure}[t]
\includegraphics[width=3in]{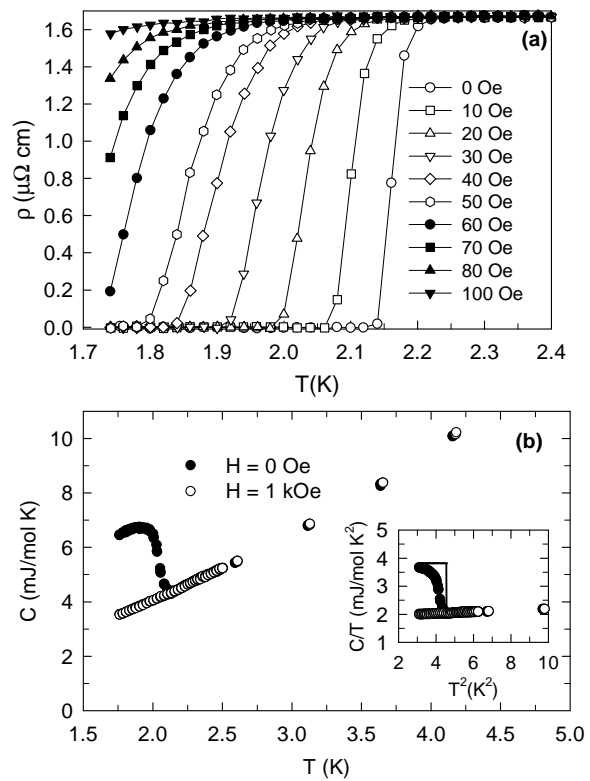}
\caption{(a) The resistivity $\rho(T)$ of OsB$_2$ between 1.7~K and 2.4~K measured with various applied magnetic fields.  (b) Temperature dependence of the heat capacity $C(T)$ of OsB$_2$ in zero and 1~kOe applied magnetic field. The inset shows the data plotted as \emph{C/T} versus $T^2$.  The solid line in a construction to estimate the magnitude of the superconducting anomaly (see text for details). 
\label{FigSC-RES-HC-OsB2}}
\end{figure}

The results of the resistivity $\rho(T)$ and heat capacity $C(T)$ measurements at low temperatures are shown in Fig.~\ref{FigSC-RES-HC-OsB2} to highlight the superconducting transition.
Figure~\ref{FigSC-RES-HC-OsB2}(a) shows the low temperature resistivity $\rho(T)$ data measured with various applied magnetic fields.  In the zero field data there is an abrupt drop below 2.2~K and $\rho$ reaches zero at $T_{\rm c}$~=~2.14~K.  The transition is quite sharp with a transition width (10\% to 90\%) of approximately 40~mK\@.  The superconducting transition is suppressed to lower temperatures in a magnetic field as can be seen in Fig.~\ref{FigSC-RES-HC-OsB2}(a).  We will return to these data when we estimate the critical field. 

Figure~\ref{FigSC-RES-HC-OsB2}(b) shows the $C(T)$ data measured in zero and 1~kOe applied field.  The sharp anomaly at 2.1~K seen in the heat capacity data taken in zero applied field confirms the bulk nature of the superconductivity in OsB$_2$.  The transition is completely suppressed to below 1.7~K in 1~kOe as seen in Fig.~\ref{FigSC-RES-HC-OsB2}(b).  The inset in Fig.~\ref{FigSC-RES-HC-OsB2}(b) shows the data plotted as \emph{C/T} versus $T^2$.  The jump in the specific heat $\Delta C$ at the superconducting transition $T_{\rm c}$ is usually normalized as $\Delta C/\gamma T_{\rm c}$ where $\gamma$ is the Sommerfeld coefficient.  Since the anomaly at the superconducting transition is somewhat broad and the temperature range below $T_{\rm c}$ is insufficient, we estimate $\Delta C$ by constructing a transition at $T_{\rm c}$ with a $\Delta C$ equal to the maximum of the superconducting anomaly in $C/T$.  This is shown as the solid vertical line in the inset of Fig.~\ref{FigSC-RES-HC-OsB2}(b).  Using this value of $\Delta C/T_{\rm c}$~=~1.78~mJ/{\rm mol}~K$^2$ and $\gamma$~=~1.90(1)~mJ/mol~K$^2$ obtained in the previous section we get $\Delta C/\gamma T_{\rm c}$~=~0.94.  This is reduced from the value 1.43 expected from BCS theory. The origin of this suppression is not currently known. A reduced specific heat anomaly at the superconducting transition ($\Delta C/\gamma T_{\rm c} \sim 1$) is also observed in the two-gap superconductor MgB$_2$.\cite{Budko2001} To evaluate this question in OsB$_2$ requires $C(T)$ data to lower temperatures than in Fig.~\ref{FigSC-RES-HC-OsB2}(b). However, we reiterate that the large specific heat jump at $T_{\rm c}$ demonstrates the bulk nature of the superconductivity in OsB$_2$. 

We now estimate the electron-phonon coupling constant $\lambda_{\rm ep}$, using McMillan's formula \cite{McMillan1967} which relates the superconducting transition temperature $T_{\rm c}$ to $\lambda_{\rm ep}$, the Debye temperature $\Theta_{\rm D}$, and the Coulumb repulsion constant $\mu^*$, 
\begin{equation}
T_{\rm c}~=~{\Theta_{\rm D} \over 1.45}\exp\left[-{1.04(1+\lambda_{\rm ep}) \over \lambda_{\rm ep} - \mu^*(1+0.62\lambda_{\rm ep})}\right]~,
\label{EqMcMillan1}
\end{equation}
\noindent
which can be inverted to give $\lambda_{\rm ep}$ in terms of $T_{\rm c}$, $\Theta_{\rm D}$ and $\mu^*$ as 
\begin{equation}
\lambda_{\rm ep}~=~{1.04+\mu^*\ln({\Theta_{\rm D} \over 1.45T_{\rm c}})\over (1-0.62\mu^*)\ln({\Theta_{\rm D} \over 1.45T_{\rm c}})-1.04}~.  
\label{EqMcMillan2}
\end{equation}
\noindent
From the above heat capacity measurements we had obtained $\Theta_{\rm D}$~=~550(11)~K and using $T_{\rm c}$~=~2.1~K we get $\lambda_{\rm ep}$~=~0.41 and 0.5 for $\mu^*$~=~0.10 and 0.15, respectively.  These values of $\lambda_{\rm ep}$ suggest that OsB$_2$ is a moderate-coupling superconductor ($\lambda_{\rm ep}$ for MgB$_2$ is $\approx{1}$).\cite{Pickett2003}

Having estimated $\lambda_{\rm ep}$, the density of states at the Fermi energy for both spin directions $N(\epsilon_{\rm F})$ can be estimated from the values of $\gamma$ and $\lambda_{\rm ep}$ using the relation \cite{Kittel}
\begin{equation}
\gamma~=~{\pi^2 \over 6}k_{\rm B}^2N(\epsilon_{\rm F})(1+\lambda_{\rm ep}) \equiv \gamma_0(1+\lambda_{\rm ep})~.
\label{EqDOSHC}
\end{equation}
\noindent
We find $N(\epsilon_{\rm F})$ ~=~1.14 and 1.06~states/(eV~f.u.) for $\lambda_{\rm ep}$~=~0.41 and 0.5, respectively.  These values are larger than the value estimated by band structure calculations $N(\epsilon_{\rm F})\approx 0.55~$(states/eV~f.u.). \cite{Hebbachea2006}  The bare Sommerfeld coefficient with $\lambda_{\rm ep}$~=~0.5 is $\gamma_0$~=~1.4~mJ/mol K$^2$.

One can now go back and re-evaluate the Wilson ratio $R_{\rm W}$.  For a free electron Fermi gas $R_{\rm W} = 1$.  In Sec.~{\ref{sec:normalstate}}, using the experimentally observed values of $\chi_{\rm P}$ and $\gamma$ we had estimated $R_{\rm W} = 1.03$ for OsB$_2$.  However, the electron-phonon interaction leads to an enhancement in $\gamma$ from its value $\gamma_0$ in the absence of interactions given by $\gamma = \gamma_0(1+\lambda_{\rm ep})$.  Similarly electron-electron interactions lead to an enhancement in the Pauli susceptibility $\chi_{\rm P}$ from its value $\chi_{\rm P}^0$ in the absence of interactions, given by $\chi_{\rm P} = {\chi_{\rm P}^0\over 1-\alpha}$, where $\alpha$ is the Stoner factor.  The re-evaluated Wilson ratio is then given by 
\begin{equation}
R_{\rm W}~=~{\pi^2 k_{\rm B}^2 \over 3\mu_{\rm B}^2}\bigg({\chi_{\rm P}^0 \over \gamma_0}\bigg)~=~1~=~{\pi^2 k_{\rm B}^2 \over 3\mu_{\rm B}^2}\bigg({\chi_{\rm P} \over \gamma}\bigg)(1-\alpha)(1+\lambda_{\rm ep})~.
\label{EqWilson1}
\end{equation}
\noindent
Using $\lambda_{\rm ep}$~=~0.41--0.5 obtained in Sec.~{\ref{sec:normalstate}}, one gets an estimate of the Stoner factor $\alpha$~=~0.30--0.35\@.  

 \label{sec:RES-SC-magnetization-acsusceptibility}
\begin{figure}[t]
\includegraphics[width=3in]{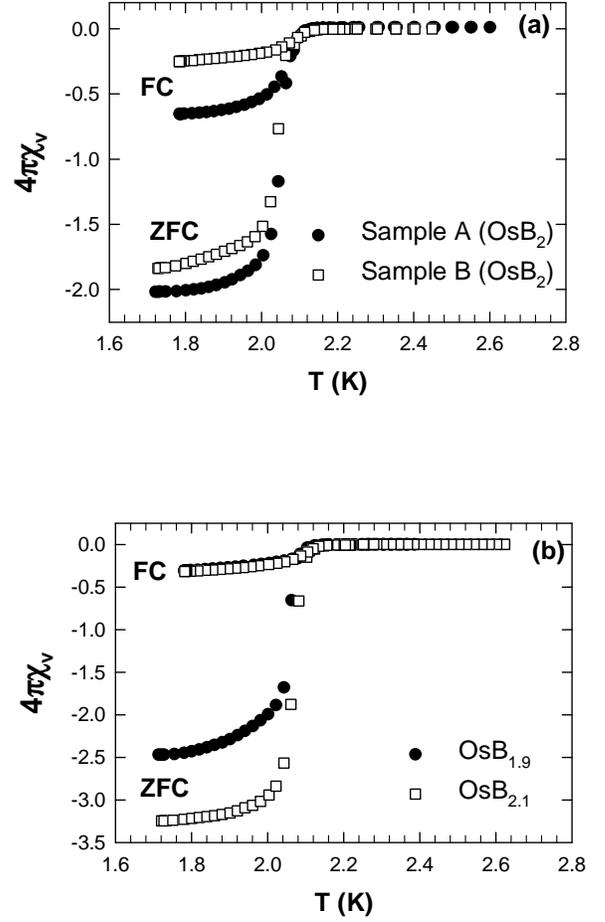}
\caption{Temperature \emph{T} dependence of the zero-field-cooled (ZFC) and field-cooled (FC) volume susceptibility $\chi_{\rm v}$ in terms of the superconducting volume fraction (4$\pi$$\chi_{\rm v}$) of OsB$_2$ (Sample~A and Sample~B) (a) and OsB$_{1.9}$ and OsB$_{2.1}$ (b) in a  field of 5~Oe from 1.7 to 2.6~K.  
\label{Figsus}}
\end{figure}

The temperature dependence of the zero-field-cooled (ZFC) and field-cooled (FC) dimensionless volume magnetic susceptibility $\chi_{\rm v}$ of OsB$_2$ samples~A and B in a field of 5~Oe from 1.7 to 2.8~K is plotted in Fig.~\ref{Figsus}(a), where $\chi_{\rm v} = M_{\rm v}/H$ and $M_{\rm v}$ is the volume magnetization.  Complete diamagnetism corresponds to $\chi_{\rm v} = -1/4\pi$, so the data have been normalized by 1/4$\pi$. The data have not been corrected for the demagnetization factor \emph{N} which gives $\chi_{\rm v} = {-1/4\pi\over 1-N}$ for the measured value.  A sharp diamagnetic drop in the susceptibility below $T_{\rm c}$~=~2.14~K for both samples signals the transition into the superconducting state.  The width of the ZFC transition (10\% to 90\% of the transition) is $\approx$~95~mK for sample~A and $\approx$~80~mK for sample~B. 

In Fig.~\ref{Figsus}(b) the temperature dependence of the susceptibility of the samples with composition OsB$_{1.9}$ and OsB$_{2.1}$ is shown.  It can be seen that the onset temperature for the superconducting transition for both the samples is 2.1~K\@.  This indicates that the homogeneity range of OsB$_2$, if any, does not have any significant effect on the $T_{\rm c}$ of OsB$_2$.  All other measurements were therefore done on the single phase samples A and B of OsB$_2$.

\begin{figure}[t]
\includegraphics[width=3.25in]{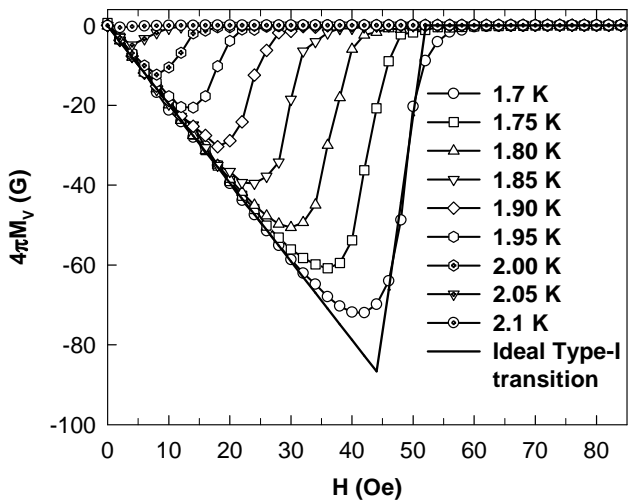}
\caption{Volume Magnetization ($M_v$) normalized by 1/4$\pi$, versus applied magnetic field (\emph{H}) at various temperatures.
\label{FigMH}}
\end{figure}

\begin{figure}[t]
\includegraphics[width=3.2in]{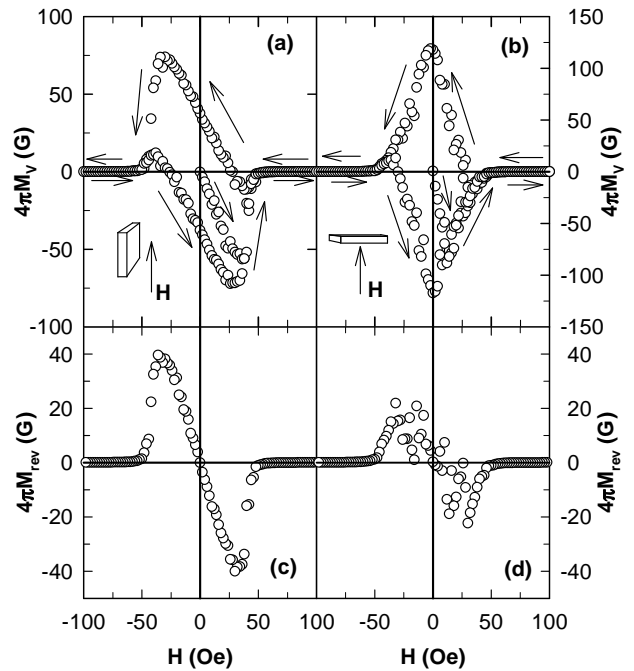}
\caption{(a) Hysteresis loop of the volume magnetization $M_{\rm v}(H)$ normalized by 1/4$\pi$, versus applied magnetic field \emph{H} at 1.7~K\@.  The magnetic field is applied parallel to the length of the sample as shown in the inset.  The arrows next to the data indicate the direction of field ramping during the measurement.  (b) Hysteresis loop of the volume magnetization $M_{\rm v}(H)$ normalized by 1/4$\pi$, versus applied magnetic field \emph{H} at 1.7~K\@.  The magnetic field is applied perpendicular to the length of the sample as shown in the inset.  (c) The normalized reversible magnetization $M_{\rm rev}$ for the data shown in (a).  (d) The normalized reversible magnetization $M_{\rm rev}$ for the data shown in (b).
\label{Fig5quadMH}}
\end{figure}

To further characterize the superconducting state we have performed measurements of the dc magnetization versus field $M(H)$ at various temperatures.  The volume magnetization $M_{\rm v}(H)$ normalized by 1/4$\pi$ is shown in Fig.~\ref{FigMH}.  The initial slope of the $M_{\rm v}(H)$ curves is larger than the value $-1$ expected for perfect diamagnetism, which indicates non-zero demagnetization effects.  The shape of the $M_{\rm v}(H)$ curves in Fig.~\ref{FigMH} for OsB$_2$ are suggestive of Type-I superconductivity with demagnetization effects.  The hysteretic $M_{\rm v}(H)$ loops, shown in Figs.~\ref{Fig5quadMH}(a) and (b), were measured at 1.75~K with the magnetic field applied parallel and perpendicular, respectively to the length of the sample as shown in the insets in Figs.~\ref{Fig5quadMH}(a) and (b).  The sample is a parallelopiped with dimensions length~=~3.25~mm, width~=~1.57~mm and thickness~=~0.35~mm.  There is a large irreversibility due to strong pinning in both measurements.  We obtain an estimate of the reversible part of the magnetization $M_{\rm rev}$ by taking an average of the magnetization measured upon increasing and decreasing $H$ between $-$100~Oe and 100~Oe.  The normalized $M_{\rm rev}$ obtained from the data in Fig.~\ref{Fig5quadMH}(a) is shown in \ref{Fig5quadMH}(c) and $M_{\rm rev}$ obtained from the data in Fig.~\ref{Fig5quadMH}(b) is shown in \ref{Fig5quadMH}(d).  The data in Fig.~\ref{Fig5quadMH}(c) look like reversible magnetization curves for a Type-I superconductor with demagnetization effects, similar to the data in Fig.~\ref{FigMH} above.  The data in Fig.~\ref{Fig5quadMH}(d) are too noisy to draw conclusions.  However, additional measurements and analysis suggest instead that OsB$_2$ is a Type-II superconductor (see below), with a Ginzburg-Landau parameter $\kappa$ on the Type-II side of the borderline between Type-I and Type-II superconductivity.  The $M_{\rm v}(H)$ in Fig.~\ref{FigMH} and Fig.~\ref{Fig5quadMH}, are however quite different from those expected for a Type-II superconductor,\cite{Brandt2003} an issue that needs to be addressed in future work.  

From the $M_{\rm v}(H)$ curves in Fig.~\ref{FigMH} we have estimated the critical field $H_{\rm c2}(T)$ from the construction in Fig.~\ref{FigMH}, illustrated for \emph{T}~=~1.7~K\@.  The $H_{\rm c2}(T)$ has been determined by fitting a straight line to the data for a given temperature in the superconducting state and to the data in the normal state and taking the field \emph{H} at which these lines intersect as the critical field at that temperature $H_{\rm c2}(T)$.   

\begin{figure}[t]
\includegraphics[width=3 in]{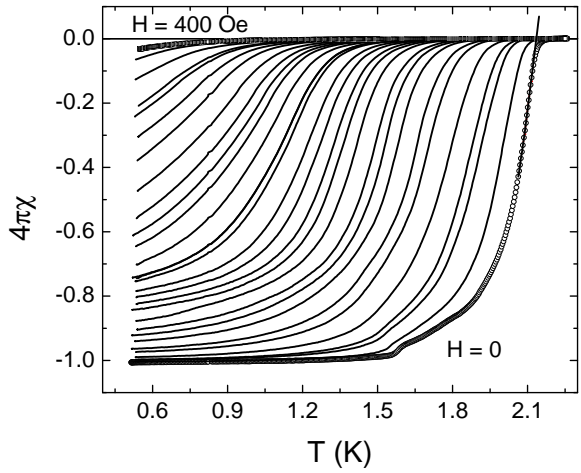}
\caption{Dynamic susceptibility $\chi$ normalized to 1/4$\pi$, versus temperature \emph{T} at a frequency of 10~MHz with various applied magnetic fields $H$.  The data have been normalized to a minimum value of $-1$ at the lowest \emph{T}.
\label{FigM(H)-vs-T}}
\end{figure}

The dynamic ac susceptibility $\chi(T)$ data measured between 0.5~K and 2.2~K at a frequency of 10~MHz in various applied magnetic fields is shown in Fig.~\ref{FigM(H)-vs-T}.  To determine $H_{\rm c2}(T)$ from the data in Fig.~\ref{FigM(H)-vs-T} we have fitted a straight line to the data in the normal state and to the data below $T_{\rm c}$ for a given applied magnetic field and taken the value of the \emph{T} at which these lines intersect as the $T_{\rm c}(H)$.  This is shown in Fig.~\ref{FigM(H)-vs-T} for the data at \emph{H}~=~0\@.  By inverting $T_{\rm c}(H)$ we obtain $H_{\rm c2}(T)$.  The $H_{\rm c2}$ has also been obtained in a similar way from the $\chi(T)\equiv M(T)/H$ SQUID magnetometer data (not shown here) between 1.7~K and 2.4~K in various applied magnetic fields.  From the $\rho(T)$ measurements [see Fig.~\ref{FigSC-RES-HC-OsB2}(b)] the applied magnetic field has been taken to be the $H_{\rm c2}$ for the temperature at which the resistance drops to zero.

 \label{sec:RES-SC-criticalfield-superfluiddensity}
\begin{figure}[t]
\includegraphics[width=3in]{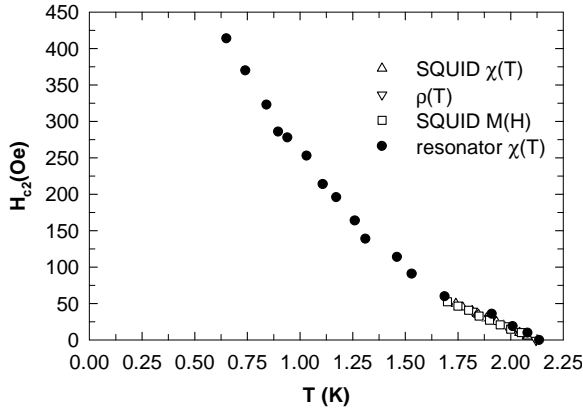}
\caption{Upper critical magnetic field $H_{\rm c2}$ versus temperature $T$ extracted from four different types of measurements.
\label{Figcrit_field}}
\end{figure}
The data for $H_{\rm c2}(T)$ obtained from all the measurements are plotted in Fig.~\ref{Figcrit_field}.  In the temperature range of the SQUID magnetometer measurements (1.7~K to 2.4~K) all the data match well and the temperature dependence of $H_{\rm c2}$ is linear.  Close to $T_{\rm c}$ the slope of the $H_{\rm c2}(T)$ curve is ${dH_{\rm c2}(T)\over dT}\sim$~$-$200~Oe/K\@.  However, there is an upward curvature at lower temperatures as seen in the $H_{\rm c2}(T)$ data extracted from the ac $\chi(T)$ measurements.  This upward curvature is sometimes seen in unconventional superconductors\cite{Zavaritsky2002, Maple1997} and also in the multi-band superconductor MgB$_2$.\cite{Shi2003}  

\begin{figure}[t]
\includegraphics[width=3in,]{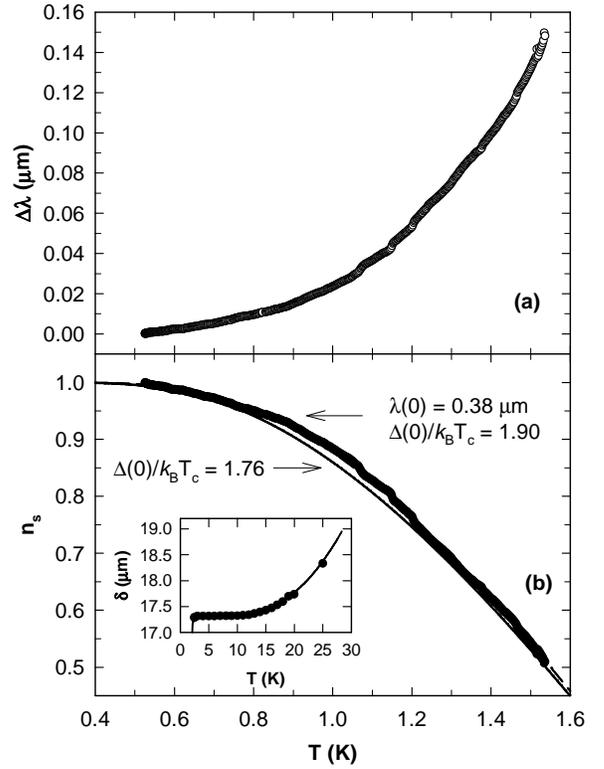}
\caption{(a) The temperature $T$ dependence of the change $\Delta\lambda$ in penetration depth in zero magnetic field $\Delta\lambda(T)\equiv \lambda(T)-\lambda(0.52~{\rm K})$. (b) The temperature dependence of the superfluid density $n_{\rm s}(T)$ of OsB$_2$.  The solid circles are the data, the dashed line is a fit by the standard \emph{s}-wave BCS model in the clean limit where the superconducting gap $\Delta(0)$ was allowed to vary and the solid line is a fit by the standard \emph{s}-wave BCS model in the clean limit with the gap fixed to the weak-coupling BCS value $\Delta(0)$~=~1.76~$k_{\rm B}T_{\rm c}$\@.  The inset shows the calibrated normal-state skin-depth (solid curve) compared to the data from resistivity measurements (solid circles).  
\label{Figsuperfluiddensity}}
\end{figure}

The temperature dependence of the change in London penetration depth, $\Delta\lambda(T)$~=~$\lambda(T) - \lambda(T_{\rm min})$~=~$\lambda(T) - \lambda(0.52~{\rm K})$, is obtained from the $\chi(T)$ data at $H = 0$ in Fig.~\ref{FigM(H)-vs-T} using Eq.~(\ref{resonatorresponse}).  The result is shown in Fig.~\ref{Figsuperfluiddensity}(a).
The temperature dependence of the superfluid density $n_{\rm s}(T)$ (the fraction of condensed electrons) can be obtained from the London penetration depth $\lambda$ using the relation \cite{Casimir1934}
\begin{equation}
n_{\rm s} = \left[{\lambda(0)\over \lambda(T)}\right]^2 ~.
\label{Eqsuperfluiddensity}
\end{equation}
The $n_{\rm s}$ can be written in terms of the measured change in London penetration depth $\Delta\lambda(T)$ as
\begin{equation}
\Delta\lambda(T)  = \lambda(T) - \lambda(0.52~{\rm K}) = \lambda(0)\Big({1\over \sqrt{n_{\rm s}}} - C \Big) ~,
\label{Eqsuperfluiddensity2}
\end{equation}
where $C$~=~$\lambda$(0.52~K)/$\lambda(0)$.  
The superfluid density depends on the shape of the Fermi surface and the superconducting gap symmetry.  If we assume a spherical Fermi surface and $s$-wave superconductivity for OsB$_2$, $n_{\rm s}$ is just a function of temperature and the magnitude of the zero temperature superconducting gap $\Delta(0)$.\cite{prozorov2006r}  We fit the $\Delta\lambda(T)$ data shown in Fig.~\ref{Figsuperfluiddensity}(a) using Eq.~(\ref{Eqsuperfluiddensity2}) with $\lambda(0)$, $C$ and $\Delta(0)$ the fitting parameters where we use $T_{\rm c} \equiv 2.15~K$ and the clean limit BCS expression for $n_{\rm s}(T,\Delta(0))$ within a semi-classical approximation.\cite{Chandrasekhar1993, prozorov2006}   The fit gave the value $\lambda(0) = 0.38(2)~\mu$m, $C$~=~1.003(2), and a slightly enhanced value for the superconducting gap $\Delta(0) = 1.90(5)~k_{\rm B}T_{\rm c}$.  Allowing $T_{\rm c}$ to vary in the fit gave no significant change to the above initially assumed value of $T_{\rm c}$.  
The $n_{\rm s}(T)$ data obtained from the estimated value of $\lambda(0)$ and the measured $\Delta\lambda(T)$ using Eq.~(\ref{Eqsuperfluiddensity}) are shown in Fig.~\ref{Figsuperfluiddensity}(b).       Figure~\ref{Figsuperfluiddensity}(b) also shows the results of the full-temperature BCS calculations for a weak-coupling \emph{s}-wave BCS model with a fixed gap $\Delta(0) = 1.76~k_{\rm B}T_{\rm c}$ (solid curve) and the result for an \emph{s}-wave BCS model with the above gap $\Delta(0) = 1.90(5)~k_{\rm B}T_{\rm c}$ (dashed curve).  The observed behaviors of $\lambda(T)$ and $n_{\rm s}(T)$ are consistent with \emph{s}-wave superconductivity.
 
We now give a thermodynamic argument in favor of Type-II superconductivity.  The zero temperature thermodynamic critical field $H_{\rm c}(0)$ of a superconductor is related to the zero temperature superconducting gap $\Delta(0)$ by the weak-coupling BCS expression\cite{Parks}
\begin{equation}
{H_{\rm c}(0)^2\over 8\pi} = {N(\epsilon_{\rm F})(1+\lambda_{\rm ep})\Delta(0)^2\over 4}~,
\label{Eqcondensationenergy}
\end{equation}
where $N(\epsilon_{\rm F})$ is the density of states at the Fermi energy for both spin directions.  The factor $(1+\lambda_{\rm ep})$ has been included on the right hand side of eq.~(\ref{Eqcondensationenergy}) to include effects of the renormalization of $N(\epsilon_{\rm F})$ due to the electron-phonon coupling.  For OsB$_2$, from our experimental $\gamma$ value and Eq.~(\ref{EqDOSHC}) one has $N(\epsilon_{\rm F})(1+\lambda_{\rm ep}) = 1.63$ states/eV f.u.~=~3.7$\times 10^{34}$~states/erg~cm$^3$\@.  Also, our penetration depth data gave $\Delta(0)/k_{\rm B}T_{\rm c}$~=~$1.90$ which gives $\Delta(0)$~=~5.6$\times 10^{-16}$~erg.  Therefore, Eq.~(\ref{Eqcondensationenergy}) gives $H_{\rm c}(0)$~=~270~G\@.  However, from Fig.~\ref{Figcrit_field} the $T$~=~0~K critical field is $\geq 420$~Oe\@.  Therefore the critical field shown in Fig.~\ref{Figcrit_field} cannot be $H_{\rm c}$, it must be $H_{\rm c2}$, and therefore OsB$_2$ is a Type-II superconductor.  The effect of stronger coupling is to reduce the calculated condensation energy\cite{Parks} and give a smaller thermodynamic critical field and so the argument above in favor of Type-II superconductivity in OsB$_2$ holds regardless.

One can estimate the Ginzburg-Landau parameter $\kappa$ using the relation\cite{Tinkham}
\begin{equation}
H_{\rm c2} = \sqrt2 \kappa H_{\rm c}~.
\label{Eqkappa0}
\end{equation}
With the value $H_{\rm c}(0)$~=~270~G obtained above and the value $H_{\rm c2}(0)\geq H_{\rm c2}$(0.52~K)~=~420~G\@, we get $\kappa(0)\geq$~1.08\@.  This value of $\kappa$ supports our above thermodynamic argument and puts OsB$_2$ on the Type-II side of the borderline ($\kappa = {1\over \sqrt{2}}$) between Type-I and Type-II superconductivity. 

\begin{figure}[t]
\includegraphics[width=3in,]{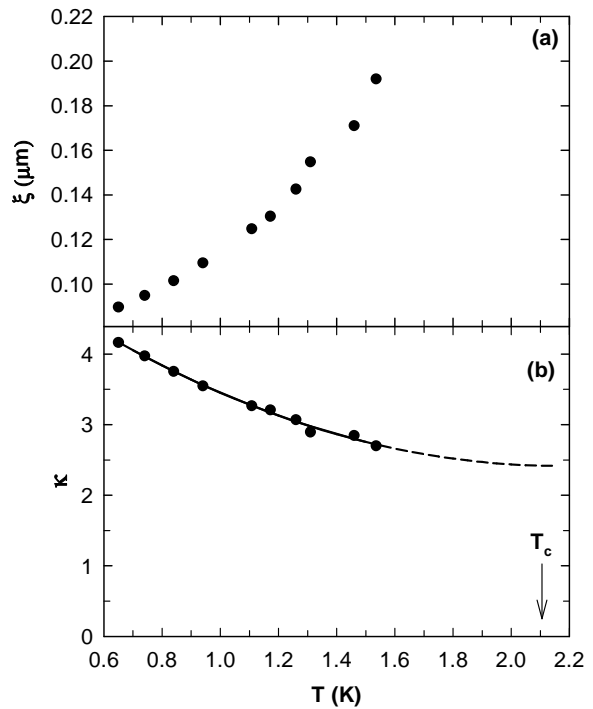}
\caption{(a) The temperature $T$ dependence of the coherence length $\xi$ and (b) of the Ginzburg-Landau parameter $\kappa$.  The solid curve in (b) is a fit by a second order polynomial temperature dependence.  The dashed line is an extrapolation up to $T_{\rm c}$ which is shown as the vertical arrow.  
\label{Figkappa}}
\end{figure}

The $\kappa$ can also be estimated from the relation\cite{Tinkham}
\begin{equation}
\kappa = {\lambda\over \xi}~.
\label{Eqkappa1}
\end{equation} 
For a Type-II superconductor, the coherence length $\xi$ can be estimated from the measured $H_{\rm c2}$ using the Ginzburg-Landau relation\cite{Tinkham} 
\begin{equation}
H_{\rm c2} = \phi_0/2\pi\xi^2   
\label{Eqkappa2}
\end{equation}
where $\phi_0$~=~$hc/2e$~=~2.068$\times 10^{-7}$~G~cm$^2$ is the flux quantum.  Since we have $H_{\rm c2}(T)$, we can get $\xi(T)$ using Eq.~(\ref{Eqkappa2}) and hence $\kappa(T) = {\lambda(T)\over\xi(T)}$ from the measured $\lambda(T)$\@.  The $\xi(T)$ and $\kappa(T)$ thus determined are shown in Fig.~\ref{Figkappa}.  Since the Ginzburg-Landau equations are written assuming a small order parameter, which means near $T_{\rm c}$,\cite{Tinkham} the dividing line between Type-I and Type-II superconductivity being 1/$\sqrt{2}$ is defined near $T_{\rm c}$.  Therefore our  $\kappa(T)$ was fitted by a second order polynomial temperature dependence to get the extrapolated value $\kappa(T_{\rm c}$~=~2.14~K)~=~2.41(3)\@.  
This value of $\kappa$ is larger compared to the estimate made above but supports the Type-II nature of superconductivity concluded above.  It should be kept in mind that the above analysis was carried out assuming isotropic superconducting properties.  Therefore, our values of $\lambda$, $\xi$ and $\kappa$ for polycrystalline samples may need revision when single crystal measurements become available.  In particular, the value of $\kappa(0.6~K)\approx$~4.2 in Fig.~\ref{Figkappa} estimated using Eq.~(\ref{Eqkappa1}) and $H_{\rm c2}(0.6~K)\approx$~420~Oe from Fig.~\ref{Figcrit_field} lead to $H_{\rm c}(0.6~K)\approx$~75~Oe from Eq.~(\ref{Eqkappa0}) which is inconsistent with the value $H_{\rm c}(0)$~=~270~Oe estimated above from Eq.~(\ref{Eqcondensationenergy}).  This inconsistency may be associated with anisotropy effects.

If the mean free path \emph{l} could be made larger by improving the quality of the sample, OsB$_2$ has the potential to be a non-elemental Type-I superconductor.  To this end we have annealed a part of sample A and performed measurements on it.  In the next section we present our results on the annealed sample.

\subsection{Annealed OsB$_2$ sample}
\label{sec:annealed}
\begin{figure}[t]
\includegraphics[width=3in]{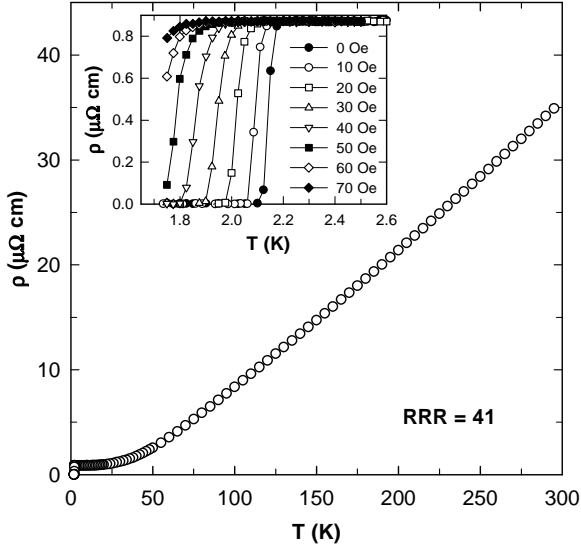}
\caption{Electrical resistivity $\rho$ versus temperature \emph{T} for annealed OsB$_2$. The inset shows the low temperature data measured in various applied magnetic fields.  
\label{FigANOsB2RES}}
\end{figure}

The $\rho(T)$ data between 1.7~K and 300~K is shown in Fig.~\ref{FigANOsB2RES}.  The room temperature value  $\rho(300~K)$~=~$36(3)~\mu$$\Omega$~cm is the same as the value observed for the unannealed sample.  The low temperature residual resistivity $\rho_0$~=~$0.87(8)~\mu$$\Omega$~cm (see inset in Fig.~\ref{FigANOsB2RES}) is smaller than the value $1.7(2)~\mu$$\Omega$~cm observed for the unannealed sample in the inset of Fig.~\ref{Fig(Os,Ru)B2RES}.  The residual resistivity ratio is 41 compared to the value 22 for the unannealed sample.  This indicates an improved sample quality with smaller amount of disorder.  The inset in Fig.~\ref{FigANOsB2RES} shows the $\rho(T)$ data between 1.7~K and 2.5~K measured with various applied magnetic fields.  The abrupt drop to zero resistance at $T_{\rm c}$~=~2.12~K seen for the zero field data is suppressed to lower temperatures with increasing field.  
\begin{figure}[t]
\includegraphics[width=3in]{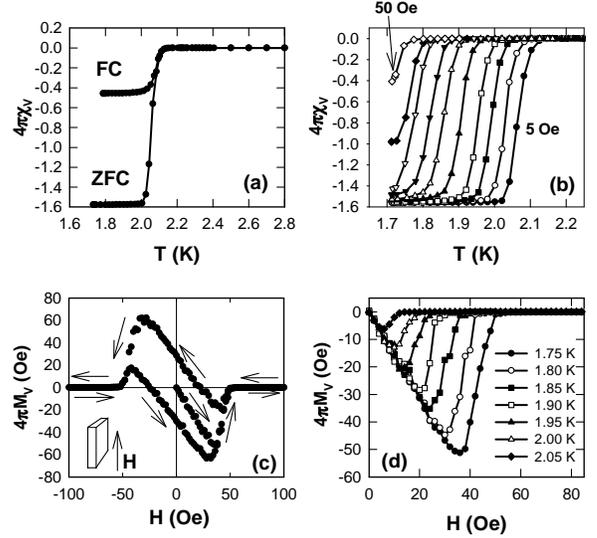}
\caption{(a) Temperature dependence of the zero-field-cooled (ZFC) and field-cooled (FC) volume susceptibility $\chi_{\rm v}(T)$ of annealed OsB$_2$, in a 5~Oe applied magnetic field, in terms of the volume fraction (4$\pi\chi_{\rm v}$).  (b) The temperature dependence of the zero-field-cooled volume susceptibility 4$\pi\chi_{\rm v}(T)$ measured in magnetic fields of 5~Oe to 50~Oe in 5~Oe steps.  (c)  Hysteresis loop of the volume magnetization $M_{\rm v}$ normalized by 1/4$\pi$, versus applied magnetic field $H$ at 1.75~K.  The magnetic field is applied parallel to the length of the sample as shown in the inset.  The arrows next to the data show the direction of field ramping during the measurement.  (d)  Volume magnetization $M_{\rm v}$ versus applied magnetic field $H$ at various temperatures. 
\label{FigAN-OsB2-chi}}
\end{figure}

The results of our magnetic measurements on the annealed OsB$_2$ sample are summarized in Fig.~\ref{FigAN-OsB2-chi}.  In the $\chi_{\rm v}(T)$ data shown in Fig.~\ref{FigAN-OsB2-chi}(a) a sharp step-like transition is seen at $T_{\rm c}$~=~2.1~K compared to the relatively rounded transition for the unannealed sample in Fig.~\ref{Figsus}.  The Meissner fraction (FC data) has also increased in the annealed sample.  Figure~\ref{FigAN-OsB2-chi}(b) shows the ZFC $\chi_{\rm v}(T)$ data between 1.7~K and 2.25~K measured in magnetic fields of 5~Oe to 50~Oe in steps of 5~Oe\@.  The superconducting transition is gradually suppressed to lower temperatures on the application of higher magnetic fields.  Figure~\ref{FigAN-OsB2-chi}(c) shows the hysteresis loop of the volume magnetization $M_{\rm v}$ versus magnetic field $H$ at 1.75~K\@.   The reversible part of the magnetization seems to have slightly increased as compared to the unannealed sample (see Fig.~\ref{Fig5quadMH}).  Figure~\ref{FigAN-OsB2-chi}(d) shows the $M_{\rm v}(H)$ data measured at various temperatures.   The data are similar to those obtained for the unannealed sample.

\begin{figure}[t]
\includegraphics[width=3in]{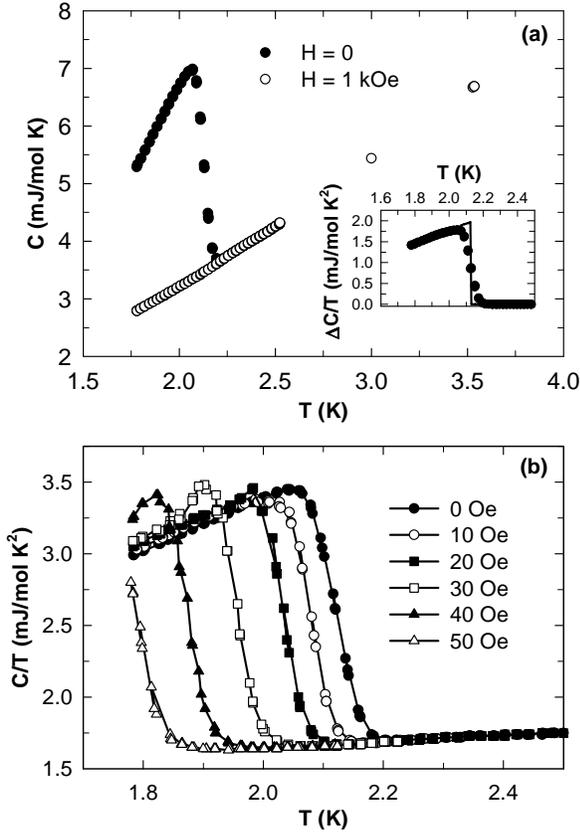}
\caption{(a) Temperature dependence of the heat capacity $C(T)$ of annealed OsB$_2$ in zero and 1~kOe applied magnetic field. The inset shows the difference $\Delta C(T)$ between the two data plotted as $\Delta C/T$ versus $T$.  The solid line in the inset is an equal entropy construction to estimate the magnitude of the superconducting anomaly (see text for details).  (b) $C/T$ versus $T$ between 1.75~K and 2.5~K measured with various applied magnetic fields.  
\label{FigAN-OsB2-HC}}
\end{figure}

The specific heat $C(T)$ data between 1.75~K and 4~K measured in zero and 1~kOe applied magnetic field are shown in Fig.~\ref{FigAN-OsB2-HC}(a).  A sharp step-like transition is seen at $T_{\rm c}$~=~2.13~K compared to the broad peak for the unannealed sample (see Fig.~\ref{FigSC-RES-HC-OsB2}).  The 1~kOe data could be fit by the expression $C(T)$~=~$\gamma T$+$\beta T^3$.  The fit gave the values $\gamma~=~1.47(3)~ {\rm mJ/mol~K}^2$ and $\beta~=~0.035(1)~{\rm mJ/mol~K}^4$.  Using the above value of $\gamma$ one can estimate the density of states for both spin directions $N(\epsilon_{\rm F})$ from Eq.~(\ref{EqDOSHC}).  One obtains $N(\epsilon_{\rm F})$~=~0.88 and 0.82~states/(eV~f.u.) for $\lambda_{\rm ep}$~=~0.41 and 0.5, respectively.  These values are somewhat smaller than the above values obtained for the unannealed sample (1.14 and 1.06~states/(eV~f.u.) respectively) but closer to the band structure value of 0.55~states/(eV~f.u.)\@.\cite{Hebbachea2006}       

The inset in Fig.~\ref{FigAN-OsB2-HC}(a) shows the difference $\Delta C(T)$ between the zero field and 1~kOe data plotted as $\Delta C(T)/T$ versus $T$.  From an equal entropy construction shown as the solid line in the inset one gets $\Delta C/\gamma T_{\rm c}$~=~1.38 which is close to the weak coupling BCS value of 1.43\@.  The Fig.~\ref{FigAN-OsB2-HC}(b) shows the specific heat data measured with various applied magnetic fields.  The transition occurs at lower temperatures with increasing field.  For magnetic fields above 20~Oe the specific heat peaks at a larger value compared to the 0~Oe and 10~Oe data before coming down to meet the zero field data at lower temperatures.   Measurements down to lower temperatures are required to see if this is a first order transition as seen in Type-I superconductors as observed for example in Ir$_2$Ga$_9$.\cite{Shibayama2007}     

\begin{figure}[t]
\includegraphics[width=3in]{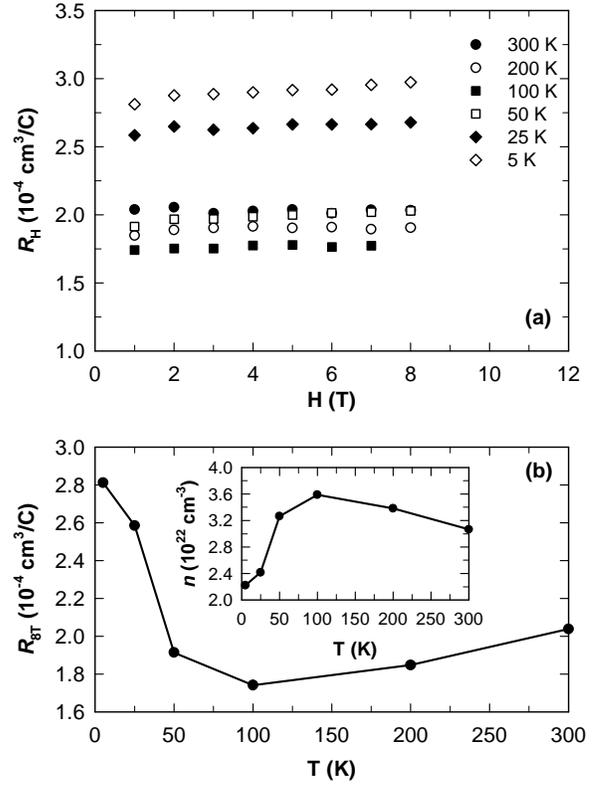}
\caption{(a) The Hall coefficient $R_{\rm H}$ versus magnetic field $H$ measured at various temperatures.  (b) The Hall coefficient $R_{\rm H}$ versus temperature $T$ in a magnetic field of 8~T\@.  The inset in (b) shows the inferred variation of the carrier density $n$ with temperature $T$ in a single-band model. 
\label{FigAN-OsB2-Hall}}
\end{figure}

Figure~\ref{FigAN-OsB2-Hall} shows the results of our Hall effect measurements.  Figure~\ref{FigAN-OsB2-Hall}(a) shows the dependence of the Hall coefficient $R_{\rm H}$ on magnetic field $H$ at various temperatures.  The $R_{\rm H}$ is positive at all fields and temperatures which indicates that hole conduction dominates the electronic transport.  The $R_{\rm H}$ is approximately constant with magnetic field at all temperatures.  From $R_{\rm H}$ one obtains a lower limit estimate of the carrier density using the single-band expression \cite{hurd1972}
$$R_{\rm H} = {1\over ne}~,$$
where $n$ is the carrier density and $e$ is the charge of the electron.  Using the value $R_{\rm H}$~=~2.03$\times$10$^{-4}$~cm$^{-3}$C$^{-1}$ at 300~K and 8~T, we get $n$~=~3.08$\times$10$^{22}$~cm$^{-3}$\@.  One can get an estimate of the density of states at the Fermi energy $N(\epsilon_{\rm F})$ using the single-band relation \cite{Kittel}    
\begin{equation}
N(\epsilon_{\rm F}) = \Big[{m (3\pi^2n)^{1\over 3}\over \hbar^2\pi^2}\Big]~,
\end{equation}
where $m$ is the free electron mass and $\hbar$ is Planck's constant divided by 2$\pi$\@.  With the value of $n$ obtained above one gets $N(\epsilon_{\rm F})$~=~0.71~states/(eV~f.u.) for both spin directions which is close to the range 0.81 -- 0.88~states/(eV~f.u.) obtained above from the heat capacity for the annealed sample.  This agreement suggests that the carrier density estimated from the Hall measurements may not be far off from the actual value for the material.

Figure~\ref{FigAN-OsB2-Hall}(b) shows the plot of $R_{\rm H}$ versus temperature $T$ and the inset shows the variation of the carrier density $n$ with $T$ at a magnetic field of 8~T\@.  The $R_{\rm H}$ shows a strong temperature dependence decreasing slightly for temperatures down to 50~K before increasing strongly at lower temperatures.  In a single band model the $R_{\rm H}$ is expected to be temperature independent if the scattering rate $\tau$ is isotropic.  However, in a two-band model, the $R_{\rm H}$ could be temperature dependent if, for example, the scattering rates of the two bands have a different temperature dependence or the fractions of charge carriers in the two bands change with temperature.\cite{hurd1972}  The strong reduction in $n$ below 50~K mimics the behavior observed for the instrinsic susceptibility shown above in Fig.~\ref{Figchi_normal}.   
  
\begin{figure}[t]
\includegraphics[width=3in]{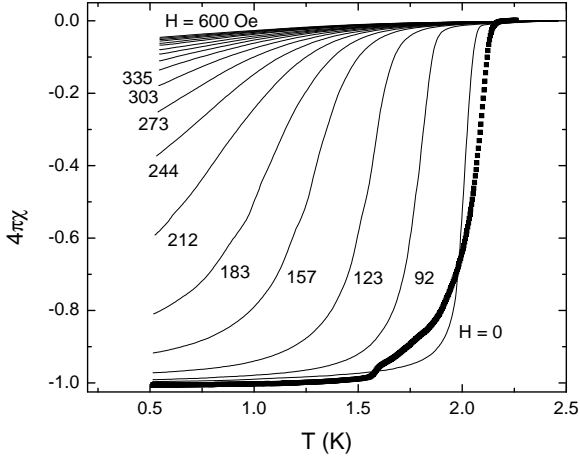}
\caption{Dynamic susceptibility $\chi$ of annealed OsB$_2$ normalized by 1/4$\pi$, versus temperature \emph{T} at a frequency of 10~MHz in various applied magnetic fields $H$.  The $H$ values in Oe are given next to the data.  The data have been normalized to a minimum value of $-1$ at the lowest \emph{T}.  For comparison the $H$~=~0~Oe data for the unannealed sample are shown as solid symbols.
\label{FigAN-M(H)-vs-T}}
\end{figure}

\begin{figure}[t]
\includegraphics[width=3in]{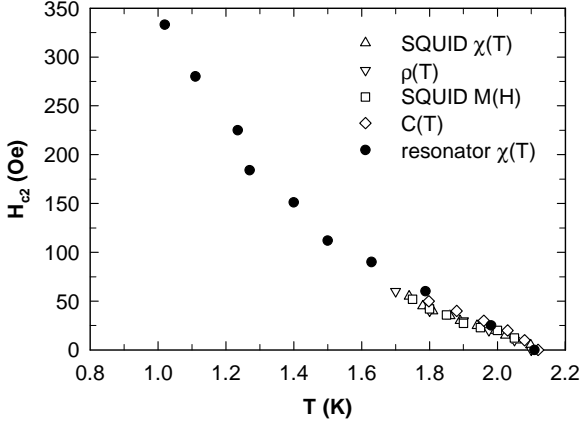}
\caption{Critical magnetic field $H_{\rm c2}$ versus temperature $T$ for annealed OsB$_2$ extracted from various measurements as noted.   
\label{FigAN-crit_field}}
\end{figure}

We now present the results of the dynamic susceptibility measurements on the annealed sample.  Figure~\ref{FigAN-M(H)-vs-T} shows the dynamic susceptibility $\chi$ of annealed OsB$_2$ normalized by 1/4$\pi$, versus temperature \emph{T} at a frequency of 10~MHz in various applied magnetic fields $H$.  The zero-field data for the unannealed sample are also shown for comparison.  The data have been normalized to a minimum value of $-1$ at the lowest \emph{T}.  The step-like feature at 1.6~K seen for the unannealed sample is no longer observed for the annealed sample suggesting an improved sample quality.  The critical field $H_{\rm c2}(T)$ was obtained from the $\chi(T)$ data shown in Fig.~\ref{FigAN-M(H)-vs-T} for the data measured in magnetic fields up to $H$~=~335~Oe for the annealed sample as described in Sec.~(\ref{sec:SC}) for the unannealed sample.  The $H_{\rm c2}(T)$ data so obtained are plotted in Fig.~\ref{FigAN-crit_field} along with the $H_{\rm c2}(T)$ data extracted from the $\chi(T)$ data in Fig.~\ref{FigAN-OsB2-chi}(b), the $\rho(T)$ data in the inset of Fig.~\ref{FigANOsB2RES}, the $M(H)$ data in Fig.~\ref{FigAN-OsB2-chi}(d) and the $C(T)$ data in Fig.~\ref{FigAN-OsB2-HC}(b).  The $H_{\rm c2}(T)$ data between 1.7~K and 2.2~K are linear with ${dH_{\rm c2}(T)\over dT}\sim$~$-$180~Oe/K near $T_{\rm c}$\@.   The data at lower temperatures show a positive curvature similar to that seen for the unannealed sample in Fig.~\ref{Figcrit_field} above.     

\begin{figure}[t]
\includegraphics[width=3in,]{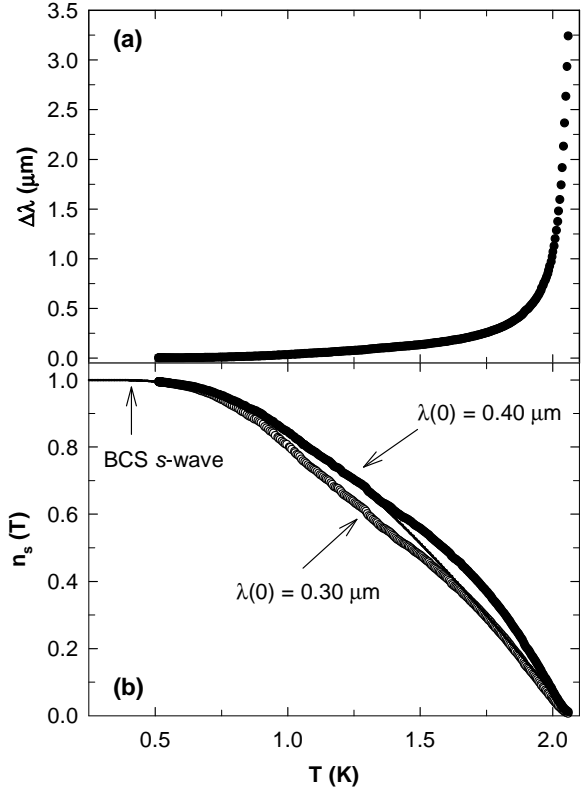}
\caption{ (a) The temperature $T$ dependence of the change $\Delta\lambda$ in penetration depth of annealed OsB$_2$ in zero magnetic field $\Delta\lambda(T)\equiv \lambda(T)-\lambda(0.52~{\rm K})$.  (b) The temperature dependence of the superfluid density $n_{\rm s}(T)$ of annealed OsB$_2$.  The filled circles are the data with $\lambda(0)$~=~0.40~$\mu$m and the open circles are the data with $\lambda(0)$~=~0.30~$\mu$m.  The solid line is the standard weak-coupling \emph{s}-wave BCS model in the clean limit.   
\label{FigAN-superfluiddensity}}
\end{figure}

The temperature dependence of the change in London penetration depth, $\Delta\lambda(T)$~=~$\lambda(T) - \lambda(T_{\rm min})$~=~$\lambda(T) - \lambda(0.52~{\rm K})$, is obtained from the $\chi(T)$ data at $H = 0$ in Fig.~\ref{FigAN-M(H)-vs-T} using Eq.~(\ref{resonatorresponse}).  The result is shown in Fig.~\ref{FigAN-superfluiddensity}(a).  The smoothly varying $\chi(T)$ data up to $T_{\rm c}$ allowed us to analyze the data to much higher temperatures than was possible for the unannealed sample which showed an unexpected bump at about 1.6~K\@.  As described in detail in Sec.~\ref{sec:SC}, we obtain the superfluid density $n_{\rm s}(T)$ data by fitting the $\Delta\lambda(T)$ data shown in Fig.~\ref{FigAN-superfluiddensity}(a) using Eq.~(\ref{Eqsuperfluiddensity2}) with $\lambda(0)$, $C$ and $\Delta(0)$ the fitting parameters.  

The $\Delta\lambda(T)$ data could not be fitted in the whole temperature range with the clean limit BCS model with a single $\lambda(0)$ value if we assume that there is only one gap.  
With a single weak-coupling $s$-wave BCS gap $\Delta(0)/k_{\rm B}T_{\rm c}$~=~1.76\@, fitting the data between 0.52~K and 1.3~K gave $\lambda(0)$~=~0.40~$\mu$m and fitting the data above 1.6~K gave $\lambda(0)$~=~0.30~$\mu$m.  The superfluid density $n_{\rm s}(T)$ data thus obtained for the two values of $\lambda(0)$ mentioned above are shown in Fig.~\ref{FigAN-superfluiddensity}(b) along with the prediction of the weak-coupling $s$-wave BCS model shown as the solid curve.  The $n_{\rm s}(T)$ data have an unusual temperature dependence with two bumps at about 0.8~K and 1.7~K\@.  This is very similar to the behavior of $n_{\rm s}(T)$ seen in polycrystalline and single crystalline MgB$_2$.\cite{Manzano2002}  

\begin{figure}[t]
\includegraphics[width=3in,]{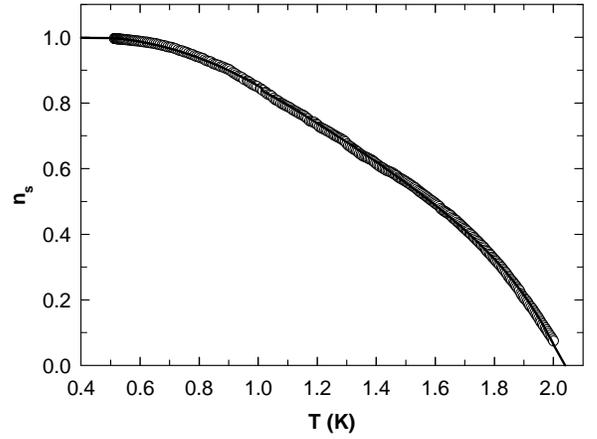}
\caption{ The temperature dependence of the superfluid density $n_{\rm s}(T)$ of annealed OsB$_2$ for $\lambda(0)$~=~0.40~$\mu$m.  The solid line is a fit by a 2-gap model in the dirty limit given by Eq.~(\ref{Eq2gap}).  
\label{FigAN-2gapsuperfluiddensity}}
\end{figure}

We were indeed able to fit our $n_{\rm s}(T)$ data with a two-gap model using the dirty limit expression\cite{gurevich2003} 
\begin{equation}
n_{\rm s}(T) = a{\Delta_1(T)\tanh({\Delta_1(T)\over 2k_{\rm B}T_{\rm c}})\over \Delta_1(0)} + b{\Delta_2(T)\tanh({\Delta_2(T)\over 2k_{\rm B}T_{\rm c}})\over \Delta_2(0)}~,
\label{Eq2gap}
\end{equation} 
\noindent
where, $\Delta_1(T)$ and $\Delta_2(T)$ are the temperature dependent gaps, $\Delta_1(0)$ and $\Delta_2(0)$ are the zero temperature values of the gaps, and $a$ and $b$ are the fractional contributions from the bands 1 and 2, respectively.  The temperature dependence of the gap is given by $\Delta(T)$~=~$\Delta(0)\tanh\big(c\sqrt{T_{\rm c}/T - 1}\big )$, where $c$ is a parameter which depends on the symmetry of the superconducting gap.\cite{poole2007}  The fit shown in Fig.~\ref{FigAN-2gapsuperfluiddensity} as the solid line through the $n_{\rm s}(T)$ data, gave the values $2\Delta_1(0)/k_{\rm B}T_{\rm c}$~=~2.8(3), $2\Delta_2(0)/k_{\rm B}T_{\rm c}$~=~3.6(4), $a$~=~0.22(3), $b$~=~0.85(4), and $c$~=~3.8(2)\@.  The two-gap model fits the data very well.  We could also fit the $n_{\rm s}(T)$ data with a two-gap model when the critical temperature $T_{\rm c}$ for the two gaps were allowed to vary. The fit (not shown here) gave a slightly smaller value for the smaller gap $2\Delta_1(0)/k_{\rm B}T_{\rm c}$~=~2.2(1) and gave a $T_{\rm c}$~=~1.5(2)~K for this gap.  The other parameters did not change much.  The similar values of $\Delta_1(0)$ and $\Delta_2(0)$ may be the reason why the specific heat jump at $T_{\rm c}$ is close to the value expected for a single weak-coupling BCS gap.  
  
We point out that for the unannealed sample, due to the step in the $\chi(T)$ at about 1.6~K the $n_{\rm s}(T)$ data in Fig.~\ref{Figsuperfluiddensity} were fitted only up to 1.5~K\@.  Therefore we do not know whether the $n_{\rm s}(T)$ data for the unannealed sample in the absence of this feature would be described by the BCS model in the whole temperature range up to $T_{\rm c}$.  In our model we assume an isotropic gap.  The deviation from the theory in Fig.~\ref{FigAN-superfluiddensity} could come from anisotropy of the gap or due to multi-band effects as in MgB$_2$.\cite{Manzano2002}  Measurements on single crystals will be needed to resolve this issue.  

Using the values $\Delta(0)$~=~1.76$k_{\rm B}T_{\rm c}$~=~5.2$\times 10^{-16}$~erg and $N(\epsilon_{\rm F}) = 0.88$ states/eV f.u.~=~2$\times 10^{34}$~states/erg~cm$^3$ obtained from the specific heat data one gets from Eq.~(\ref{Eqcondensationenergy}) $H_{\rm c}(0)$~=~190~G\@.  The critical field for the annealed sample at 1.0~K in Fig.~\ref{FigAN-crit_field} is already about 350~G\@.  Therefore annealed OsB$_2$ is still a Type-II superconductor.  An estimate of the Ginzburg-Landau parameter $\kappa$ can now be made using Eq.~(\ref{Eqkappa0}).  With the value $H_{\rm c}(0)$~=~190~G obtained above and the value $H_{\rm c2}(0)\geq H_{\rm c2}$(1~K)~=~335~G\@, we get $\kappa(0)\geq$~1.2 indicating the Type-II nature of the superconductivity of the annealed sample.   

\begin{figure}[t]
\includegraphics[width=3in,]{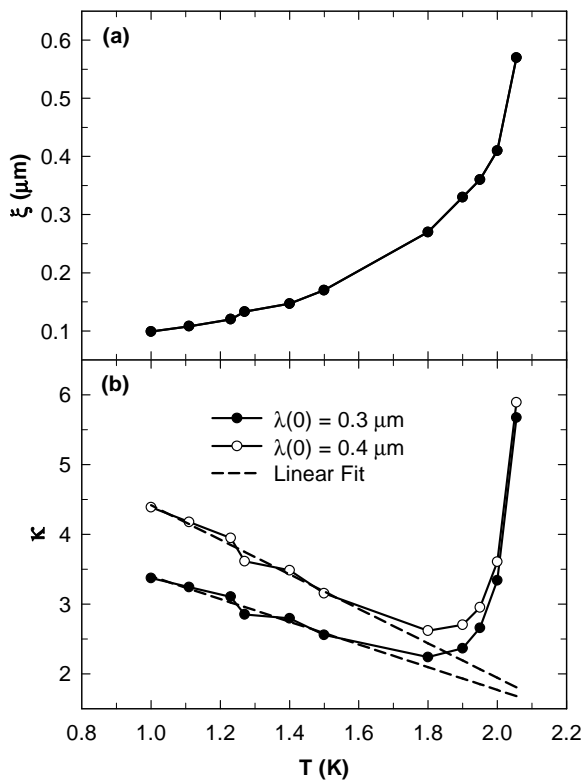}
\caption{(a) The temperature $T$ dependence of the coherence length $\xi$ for annealed OsB$_2$ and (b) of the Ginzburg-Landau parameter $\kappa$ for annealed OsB$_2$ for $\lambda(0)$~=~0.3 and 0.4~$\mu$m\@.  The solid curves through the data are guides to the eye.  The dashed curves in (b) are linear fits to the $\kappa(T)$ data between 1.0~K and 1.5~K and extrapolated to $T_{\rm c}$\@.
\label{FigAN-kappa}}
\end{figure}

From the $H_{\rm c2}(T)$ data in Fig.~\ref{FigAN-crit_field} we can get $\xi(T)$ using Eq.~(\ref{Eqkappa2}) and from Eq.~(\ref{Eqkappa1}) we can get $\kappa(T)$ using the $\Delta\lambda(T)$ data in Fig.~\ref{FigAN-superfluiddensity}(a) with $\lambda(0)$~=~0.3 or 0.4~$\mu$m obtained above.  The $\xi(T)$ and $\kappa(T)$ data thus obtained are plotted in Figs.~\ref{FigAN-kappa}(a) and (b), respectively.   The behavior of $\kappa(T)$ near $T_{\rm c}$ is unusual.  Such a behavior has been reported before for small spheres of the Type-I superconductors tin and indium,\cite{feder1969} where it was suggested that the divergence of the measured $\kappa$ near $T_{\rm c}$ comes from finite size effects; near $T_{\rm c}$ when the coherence length and penetration depth become comparable to the size of the sample, the apparent $\kappa$ diverges.  A similar increase of $\kappa$ near $T_{\rm c}$ has also been reported recently for single crystals of the Type-I superconductor Ag$_5$Pb$_2$O$_6$.\cite{Yonezawa2005}
Divergence of several quantities including $\kappa$ near $T_{\rm c}$ has also been reported in high-$T_{\rm c}$ materials and this behavior was shown to arise from the interaction of vortices near $T_{\rm c}$ where the size of the vortices became comparable to the distance between the vortices.\cite{kogan1993}  The reason for such a behavior for OsB$_2$ is unclear and further measurements will be required to understand this issue.  To get an alternative estimate of $\kappa$ near $T_{\rm c}$ we have fitted the $\kappa(T)$ data between 1.0~K and 1.5~K and extrapolated to $T_{\rm c}$.  The fits, shown as the dashed straight lines through the data in Fig.~\ref{FigAN-kappa} gave the values $\kappa(T_{\rm c})$~=~1.80 and 1.63 for $\lambda(0)$~=~0.4 and 0.3~$\mu m$, respectively.   
   
Another estimate of $\kappa$ can be made as follows.  First, the residual resistivity $\rho_0=0.87~\mu\Omega$~cm (see inset in Fig.~\ref{FigANOsB2RES}) gives the mean free path $l = mv_{\rm F}/\rho_0ne^2 = 0.084~\mu$m, where \emph{m} is taken to be the free electron mass, $v_{\rm F}$ is the Fermi velocity, $n$ is the conduction electron density and \emph{e} is the electron charge.  Using the value $n=3.08\times 10^{22}$ cm$^{-3}$ obtained from the Hall effect measurements, in a single-band model one has \cite{Kittel} $v_{\rm F}$~=~$({h\over 2\pi m})\big(3\pi^2n\big)^{2\over 3} \approx 5.9\times 10^7$~cm/s.  The BCS coherence length is then \cite{Tinkham} $\xi_0 = \hbar v_{\rm F}/\pi\Delta(0) \simeq 0.38~ \mu$m.  
Since $\xi_0$ is larger than $l$, annealed OsB$_2$ is not in the clean limit.  The effective coherence length is then \cite{Tinkham} $\xi(0)$~=~$0.85\sqrt{\xi_0l}$~=~0.151~$\mu$m and the Ginzburg-Landau parameter $\kappa$ can be estimated using Eq.~(\ref{Eqkappa1}).  
We get $\kappa(0)$~=~1.99 and 2.6 for $\lambda(0)$~=~0.30 and 0.40~$\mu$m, respectively.  
These values of $\kappa$ are consistent with the value $\kappa(0)\geq$1.2 obtained from the thermodynamic argument above, and also with the $\kappa(T)$ values obtained from linear fits to the $\kappa(T)$ data above.  These values support our conclusion that OsB$_2$ is a Type-II superconductor.

\section{CONCLUSION}
\label{sec:CON}
We have synthesized the compounds OsB$_2$ and RuB$_2$ and measured their magnetic, transport and thermal properties.  Our measurements confirm that OsB$_2$ undergoes a bulk transition into the superconducting state below 2.1~K\@.  Analysis of our data suggests that OsB$_2$ is a moderate-coupling ($\lambda_{\rm ep}$~=~0.4 to 0.5) Type-II superconductor with a small Ginzburg-Landau parameter $\kappa \sim$ 1--2 and an upper critical field $H_{\rm c2}(0.5~{\rm K})~\sim~420$~Oe for an unannealed sample and $H_{\rm c2}(1~{\rm K})~\sim~330$~Oe for an annealed sample.  The reduced specific heat jump at $T_{\rm c}$ observed for the unannealed sample and the positive curvature in $H_{\rm c2}(T)$ observed for both the unannealed and annealed samples are similar to effects observed in MgB$_2$.  The temperature dependence of the superfluid density $n_{\rm s}(T)$ is consistent with an \emph{s}-wave superconductor with a slightly enhanced gap $\Delta(0) = 1.90(5)~k_{\rm B}T_{\rm c}$ and a zero temperature London penetration depth $\lambda(0)$~=~0.38(2)~$\mu$m for the unannealed OsB$_2$ sample and $\Delta(0) = 1.76~k_{\rm B}T_{\rm c}$ and $\lambda(0)$~=~0.30--0.40~$\mu$m for the annealed OsB$_2$ sample.  For the annealed sample the $n_{\rm s}(T)$ data above 1.4~K deviate from the predictions of the $s$-wave BCS model and show a shoulder at 1.8~K\@.  This behavior is similar to the behavior observed in MgB$_2$.  In the normal state, unannealed OsB$_2$ and RuB$_2$ are Pauli paramagnetic metals with very similar properties.  To investigate the effect of boron off-stoichiometry on the superconducting properties of OsB$_2$ we also investigated samples with starting compositions OsB$_{1.9}$ and OsB$_{2.1}$.  Both samples had a transition temperature $T_{\rm c}$~=~2.1~K indicating no significant dependence of $T_{\rm c}$ on the boron stoichiometry.  The diminished specific heat jump at $T_{\rm c}$ in Fig.~\ref{FigSC-RES-HC-OsB2}(b) for the unannealed sample, the non-Type-II shape of the superconducting $M(H)$ in Fig.~\ref{FigMH} and Fig.~\ref{FigAN-OsB2-chi}(d), the positive curvature in $H_{\rm c2}(T)$ in Fig.~\ref{Figcrit_field} and Fig.~\ref{FigAN-crit_field}, and the non-BCS like temperature dependence of the $n_{\rm s}(T)$ in Fig.~\ref{FigAN-superfluiddensity}(b) are all interesting issues that require further investigation.  Future studies of the anisotropic physical properties of single crystals will be very helpful in these regards.

\begin{acknowledgments}

We thank V. G. Kogan, J. R. Clem, D. K. Finnemore and P. C. Canfield for useful discussions.  Work at the Ames Laboratory was supported by the Department of Energy-Basic Energy Sciences under Contract No.\ DE-AC02-07CH11358.  R.P. also acknowledges support from NSF Grant number DMR-05-53285 and from the Alfred P. Sloan Foundation.
\end{acknowledgments}

\end{document}